\documentstyle[epsfig]{mn}

\title[Deriving star formation histories]{Deriving star formation histories:
Inverting HR diagrams through a 
variational calculus maximum likelihood method}
\author[X. Hernandez, D. Valls-Gabaud and
G. Gilmore]{X. Hernandez$^{1,2}$, 
David Valls-Gabaud$^{3,4}$ and Gerard Gilmore$^1$ \\
$^1$ Institute of Astronomy, Cambridge University, Madingley Road, Cambridge CB3 0HA \\
$^2$ Instituto de Astronom\'\i a, Universidad Nacional Aut\'onoma de
M\'exico, A.P. 70-264, 04510 M\'exico, D.F. \\
$^3$ UMR CNRS 7550, Observatoire de Strasbourg, 11 Rue de l'Universit\'e, 67000 Strasbourg, France. \\
$^4$ Royal Greenwich Observatory, Madingley Road, Cambridge CB3 OEZ.} 

\date{\today}

\begin{document}
\maketitle

\begin{abstract}

We introduce a new method for solving maximum likelihood problems through
variational calculus, and apply it to the case of recovering an
unknown star formation history, $SFR(t)$, from a resulting HR diagram. This
approach allows a totally non-parametric solution which has the
advantage of requiring no initial assumptions on the $SFR(t)$.
As a full maximum likelihood statistical model is used, we take
advantage of all the information available in the HR diagram, rather than
concentrating on particular features such as turn off points or luminosity functions.
We test the method using a series of synthetic HR diagrams produced
from known $SFR(t)$, and find it to be quite successful under noise
conditions comparable to those present in current observations. 
At this point we restrict the analysis to situations where the metallicity of the
system is known, as is the case with the resolved populations of 
the dwarf spheroidal companions to the Milky Way or the solar neighbourhood
Hipparcos data. We also include tests
to quantify the way uncertainties in the assumed metallicity, binary fraction
and IMF affect our inferences. 

\end{abstract}


\section{Introduction}

Star formation histories, as described by the temporal evolution of the star formation
rate, $SFR(t)$, and the cumulative numbers of stars formed are the
crucial astronomical parameters needed to constrain cosmological models, 
galaxy formation scenarios, galactic evolution models and star formation theories
in general. In practice, uncertainties in the theories of stellar formation and evolution,
as well as degeneracy in a stars observational parameters between age and metallicity,
not to mention observational errors and unknown distance and reddening corrections,
make inferring $SFR(t)$ for distant systems difficult. Actually, even assuming a
known stellar initial mass function ($IMF$) and 
metallicity, a given set of isochrones and no distance or reddening
corrections, recovering the $SFR(t)$ which gave rise to a given HR diagram is not
trivial. It is this restricted form of the general problem which we attempt to improve
on in this paper. 

	The increasing feasibility of studies which resolve the stellar populations
of nearby systems has made possible quantitative investigation of the $SFR(t)$
in these systems through comparison of the observed HR diagram with synthetic ones
e.g. Chiosi et al. (1989), Aparicio et al. (1990) and Mould et al. (1997) using Magellanic
and local clusters, and Mighell \& Butcher (1992), Smecker-Hane et al. (1994), Tolstoy (1995), 
Aparicio \& Gallart (1995) and Mighell (1997)
using dSph companions to the Milky Way.
The framework within which this problem is generally faced is to construct
a statistical estimator of how closely a synthetic HR diagram constructed from an
assumed $SFR(t)$ resembles the observed one, and then to select the $SFR(t)$, from amongst
a set of plausible ones, which maximizes the value of this estimator (e.g. Tolstoy \& Saha 1996).
The most rigorous 
estimator is probably the likelihood, as defined through Bayes's theorem. In practice 
this states that one should look for the model which maximizes the probability of
the observed data set having arisen from it. In comparing two or more
candidate models through the likelihood one takes into account the
position of each star in the observed HR diagram, there being no
necessity to smooth the data into a continuous distribution, or to
include only specific features of the HR diagram, such that all the
available information contributes to the comparison. The robustness of the approach is undermined
by the degree of subjectivity associated with defining the set of plausible models one
is going to consider. Further, as none of the statistical estimators has an absolute 
normalization, in the end one is left with that model, of the ones one started by proposing,
which best reproduces the data, which might not necessarily be a ``good'' 
approximation to the true $SFR(t)$. The likelihood of the data having arisen 
from a particular model can only be calculated if one has the data, the errors, and
the particular model fully specified. This last condition has led to the almost
exclusive use of parametric $SFR(t)$'s. 

This last point is not necessarily 
negative, in the case of globular clusters both theoretical calculations showing the
binding energy of the gas in the clusters to be insufficient to retain the gas
against the disruption of the first SN's and the observed thin, clearly defined 
isochrones in the HR diagrams imply that globular clusters consist of approximately coeval
stellar populations. In this case, the $SFR(t)$ can accurately be parameterized as
a short lived burst. Given a normalization condition through the total
number of stars, the problem is reduced to finding the age of the burst i.e., 
a 1D parameter space is defined, which includes all plausible $SFR(t)$. The likelihood
is then evaluated over all the parameter space, and the value of the parameter 
for which the maximum occurs selected. However, in most other stellar systems, 
one expects a more complex $SFR(t)$, where a particular parameterization will 
sometimes appear to reject a particular astrophysical model and favour another. 
If it is the precise form of the $SFR(t)$ which might serve as a constraint
on a theory (e.g. a collection of randomly located bursts as fragments accrete or a more
uniform function as gas cools, for the build up of the Galaxy), one must 
consider the most general $SFR(t)$. The less one assumes {\it a priori} about 
the $SFR(t)$ one is solving for, the more objective the inference will be.

	A first attempt at solving for $SFR(t)$ non parametrically is to break
the star formation history into a series of bursts, and to solve for the amplitude 
of each one e.g. Dolphin (1997). Other variants of this approach are possible, 
for example Hurley-Keller et al. (1998) who parameterize the $SFR(t)$ of the 
Carina dwarf as consisting of 3 bursts, and solve for the positions, durations
and amplitudes of each. 
In principle, as the number of bursts considered tends
to infinity, the full $SFR(t)$ is recovered. The difficulty in increasing the
number of bursts considerably lies in that each extra burst increases
the {\it dimension} of the parameter space by at least one. As the likelihood hyper-surface 
will in general be quite complex, the only reliable way of finding the absolute
maximum is to evaluate the likelihood function over the entire parameter space.
This last procedure is clearly not a practical approach, as calculating the
likelihood of a complete set of thousands of observed stars for even one
single model is a lengthy procedure, let alone throughout a 100 or more
dimensional space. 

Further, methods which consider a large parameter space often do not use 
a full likelihood analysis, but simpler statistical estimators such as luminosity functions 
(Aparicio \& Gallart 1995, Mighell 1997).
In this last approach the HR diagram is divided into cells and the numbers of stars in each
used as independent variables to construct a statistical estimator. The resulting 
statistic is not strictly rigorous as the numbers of stars in different cells are in fact
correlated through the underlying IMF and $SFR(t)$. Presently, methods of
comparing simulated HR diagrams with observations can be classified according to the
statistical criterion used in the comparison. A few examples of the variety in these categories
are Tolstoy (1995) and Mould et al. (1997) who use full maximum likelihood statistics, Dolphin (1997)
and Ng (1998) who use chi-squared statistics, and Aparicio et al. (1997) and Hurley-Keller et al. (1998) 
who break the HR diagrams into luminosity functions before constructing the statistical estimator.

In this paper we present a variational calculus method of solving
directly for the maximum likelihood $SFR(t)$, which does not require
any assumptions on the function one is trying to recover, or to
evaluate the likelihood of any of the $SFR(t)$'s being considered (all
continuous functions of time). We construct an integro-differential
equation which is iterated to find a $SFR(t)$ which  yields a
vanishing first variation for the likelihood. At each iteration the $SFR(t)$ is
solved with an arbitrary time resolution. Conveniently, computation
times scale only linearly with this time resolution. This allows a
very fine reconstruction of the $SFR(t)$, which would be prohibitively
expensive in a parametric decomposition of the $SFR(t)$.

The layout of our paper is as follows: in section 2 we present the
method, which we test in section 3 using a variety of synthetic HR
diagrams. In section 4 we explore how errors in the assumed 
metallicity, IMF and binary fraction modify our results, and 
section 5 contains our conclusions.

\section{Deriving star formation histories}

As stated in the introduction, our goal is to recover the star
formation history which gave rise to an observed population of stars,
described by $SFR(t)$, the star formation rate as a function of
time. As we want our method to be of a very general applicability, we
shall assume absolutely nothing about the $SFR(t)$ we are trying to
recover, beyond that it be a continuous function of time. It is
important not to impose any {\it a priori} parameterization on
$SFR(t)$, since it is precisely the form of this function that we are
trying to recover from the data, $SFR(t)$ will be fixed entirely by
the data. One obvious constraint will be the total number of stars
produced, which furnishes a normalization condition on $SFR(t)$, over
the range of masses over which stars can be observed. Obtaining the
faint end slope of the initial mass function is an entirely different
problem which we shall not address. Here we will be concerned only
with that fraction of the total star formation which produced stars
still readily visible today. It is this that we are calling the $SFR(t)$. 

The final
observed HR diagrams are the result of the star formation histories in
those systems, but also of the relevant initial mass function, the
metallicity and the stellar evolutionary processes. We shall assume all
these other physical ingredients are actually known. This is not an
unreasonable assumption, since there are many interesting
astrophysical systems for which this essentially holds, and for which
only the star formation history is poorly known. Examples of such
systems are some of the dwarf spheroidal companions to our Galaxy, for which
independent measurements of the metallicity confirm that this parameter
is consistently low throughout (e.g. Da Costa 1994).

Further, we are only interested in the
stars which end up in the observations, which basically leaves us with
a mass region over which the initial mass function is well
established. Theoretical studies of stellar isochrones have advanced
significantly over the last decade, and now there seems to be little
uncertainty in the physical properties of stars over the mass range
0.6-3 solar masses, during all but the shortest lived periods. Here we are using
the latest Padova isochrones (Fagotto et al. 1994, Girardi et al. 1996), 
including most stages of stellar evolution up to the RGB phase. 
Our inference will depend on the precise details of the isochrones we
use, but our aim here is not to insist upon any particular age calibration, but
basically to show the merits of the method being presented. Future papers apply the
method to real data.

\subsection{The method}

Having a fixed set of observations $A=(A_1,...,A_n)$, which we are
assuming resulted from a model which belongs to a certain known set of
models $B={B_1,...}$ we want to
find the model which has the highest probability of resulting in
the observed data set, $A$. That is, we wish to identify the model which maximizes
$P(AB_i)$, the joint probability of $A$ occurring for a given model
$B_i$. From the definition of conditional probabilities,

\begin{equation}
P(AB_i)=P(A|B_i)\cdot P(B_i)=P(B_i|A) \cdot P(A)
\end{equation}

where $P(A|B_i)$ is the conditional probability of observing A given
a fixed model $B_i$ occurred, $P(B_i|A)$ is the conditional probability of
model $B_i$ given the observed data $A$, and $P(A), P(B_i)$ are
the independent probabilities of $A$ and $B_i$,
respectively. Further, if the $B_{i}$s are exclusive and exhaustive,

\begin{equation}
P(A)=\int_{i} P(A|B_i) \cdot P(B_i)=1/C
\end{equation}

where $C$ is a constant, so that equation~(1) becomes:

\begin{equation}
P(B_i|A)=C\cdot P(A|B_i)\cdot P(B_i)
\end{equation}

which is Bayes' theorem. $P(B_i)$ is called the {\it prior}
distribution, and defines what is known about model $B_i$ without
any knowledge of the data. As we want to maximize the relevance of the
data in our inference, we can take the hypothesis of equal prior
probabilities, finding the maximum likelihood model under this
assumption is hence simplified to finding the model $B_i$ for which
$P(A|B_i)$ is maximized. Our set of models from which the optimum
$SFR(t)$ is to be chosen includes all continuous, twice differentiable
functions of time such that the total number of stars formed does not 
conflict with the observed HR diagram.

In order to find the $SFR(t)$ which maximizes the probability of the
observed HR diagram resulting from it, we first have to introduce a
statistical model to calculate the probability of the data resulting
from a given $SFR(t)$. Take one particular star, having an observed
luminosity and colour, $l_{i}, c_{i}$, and an intrinsic luminosity and
colour $L_{i}, C_{i}$, which will usually differ due to observational
errors, where the index $1<i<n$
distinguishes between the $n$ observed stars making up the HR
diagram. The probability of this observed point being a star belonging
to a particular isochrone $C(L;t_j)$, i.e., being part of the stars
formed by $SFR(t_j)$ will be given by:

\begin{equation}
P_{i}\left(t_{j}\right) = SFR(t_{j}) {\rho(L_i;t_{j}) \over{\sqrt {2 \pi} \sigma(l_i)}} 
exp\left(-\left[C(L_i;t_{j})-c_{i}\right]^2 \over {2 \sigma^2(l_i)} \right) 
\end{equation}

In equation~(4) $\sigma(l_i)$ denotes the observational error in the measurement
of the colour of the $ith$ observed star, which is a function of the
luminosity of this star, and which we are assuming follows a Gaussian distribution. 
In real data, the errors in the luminosity
are much smaller than in the colour determination, which comes from
subtracting two observed quantities. For simplicity, we only consider
errors in the colour, which increase with decreasing luminosity, in a
way determined by the particular observation. In this case
$L_{i}=l_{i}$ which we adopt throughout, the generalization to an
error ellipsoid being trivial. 
$C(L_i;t_j)$ is the colour the observed star having luminosity $l_i$
would have if it had actually formed at $t=t_j$.
$\rho(L_i;t_{j})$ is the density of stars along the isochrone
$C(L;t_j)$ around the luminosity of the observed star, $l_i$, for
an isochrone containing a unit total mass of stars. Therefore, for
stars in their main sequence phase, $\rho(L;t_{j})$ is actually the
initial mass function expressed in terms of the luminosity of the
stars. Further along the isochrone it contains the initial mass
function convolved with the appropriate evolutionary track. Finally,
$SFR(t_j)$ indicates the total mass of stars contained in the
isochrone in question, and is the only quantity in equation~(4) which we
ignore, given an observational HR diagram, an initial mass function and
a continuous set of isochrones.

The probability of the observed point $l_{i}, c_{i}$ being the result
of a full given $SFR(t)$ will therefore be:

\begin{equation}
P_{i}\bigl( SFR(t)\bigr) = \int_{t_0} ^{t_1} SFR(t) G_{i}(t) dt
\end{equation}

where
$$
G_{i}(t)= {\rho(L_i;t) \over{\sqrt{2 \pi} \sigma(l_i)}} 
exp\left(-\left[C(L_i;t)-c_{i}\right]^2 \over {2 \sigma^2(l_i)} \right)
$$

where $t_0$ and $t_1$ are a maximum and a minimum time needed to be
considered, for example 0 and 15 Gyr. We shall refer to $G_{i}(t)$ as the
likelihood matrix. At this point we introduce the
hypothesis that the $n$ different observed points making up the total HR
diagram are independent events, to construct: 

\begin{equation}
{\cal L}= \prod_{i=1}^{n} \left( 
\int_{t_0} ^{t_1} SFR(t) G_{i}(t) dt \right)
\end{equation}

which is the probability that the full observed HR diagram resulted
from a given $SFR(t)$. This first part is essentially well known, and we have
presented it as it was laid out in Tolstoy \& Saha (1996), who use equation (6)
to compare between different set proposed $SFR(t)'s$. 

The remainder of the development is entirely new. We shall use
equation (6) to construct the Euler equation of the problem, and hence obtain
an integro-differential equation directly for the maximum likelihood $SFR(t)$, 
about which we shall assume nothing {\it a priori}.
It is the functional ${\cal L}(SFR(t))$ which we want to
maximize with respect to $SFR(t)$ to find the maximum likelihood star
formation history.

The condition that ${\cal L}(SFR)$ has an extremal can be written as
$$
\delta {\cal L}(SFR)=0,
$$
and the techniques of variational calculus brought to bear on the
problem. Firstly, we develop the product over $i$ using the chain
rule for the variational derivative, and divide the resulting sum by ${\cal L}$ to obtain:

\begin{equation}
\sum_{i=1}^{n} \left(
{\delta \int_{t_0} ^{t_1} SFR(t) G_{i}(t) dt} \over {\int_{t_0}^{t_1} SFR(t) G_{i}(t) dt}
\right) =0
\end{equation}

In order to construct an integro-differential equation for $SFR(t)$ we
introduce the new variable $Y(t)$ defined as:

$$
Y(t)=\int{ \sqrt {SFR(t)} dt} \Longrightarrow  SFR(t)=\left( {dY(t)
\over dt} \right)^2
$$

Introducing the above expression into
equation~(7) and developing the Euler equation yields, 

\begin{equation}
{d^2 Y(t)\over dt^2}\sum_{i=1}^{n} \left( G_{i}(t) \over I(i)\right)
=-{dY(t)\over dt}\sum_{i=1}^{n} \left( dG_{i}/dt \over I(i)\right)
\end{equation}

where 
$$
I(i)=\int_{t_0}^{t_1} SFR(t) G_{i}(t) dt
$$

We have thus constructed an integro-differential equation whose solution
yields a $SFR(t)$ for which the likelihood has a vanishing first variation.
This in effect has transformed the problem from one of searching for a function
which maximizes a product of integrals (equation 6) to one of solving
an integro-differential equation (equation 8).
Solving equation~(8) will be the main problem, as this would yield the
required star formation history directly, without having to
calculate ${\cal L}$ explicitly over the whole space containing all the
possible $SFR(t)s$. 
	We now implement an iterative scheme for solving equation(8), the
details of which are given in the appendix. Given the
complexity of the isochrones, the initial mass function and the unknown
star formation histories we are trying to recover, it is not possible
to prove convergence analytically for the implemented iterative method. We shall
show experimentally in the following section that the method works
remarkably well for a wide range of synthetic HR diagrams produced
from known $SFR(t)$'s, independent of the initialization used.

\section{Testing the method}

\begin{figure*}
\epsfig{file=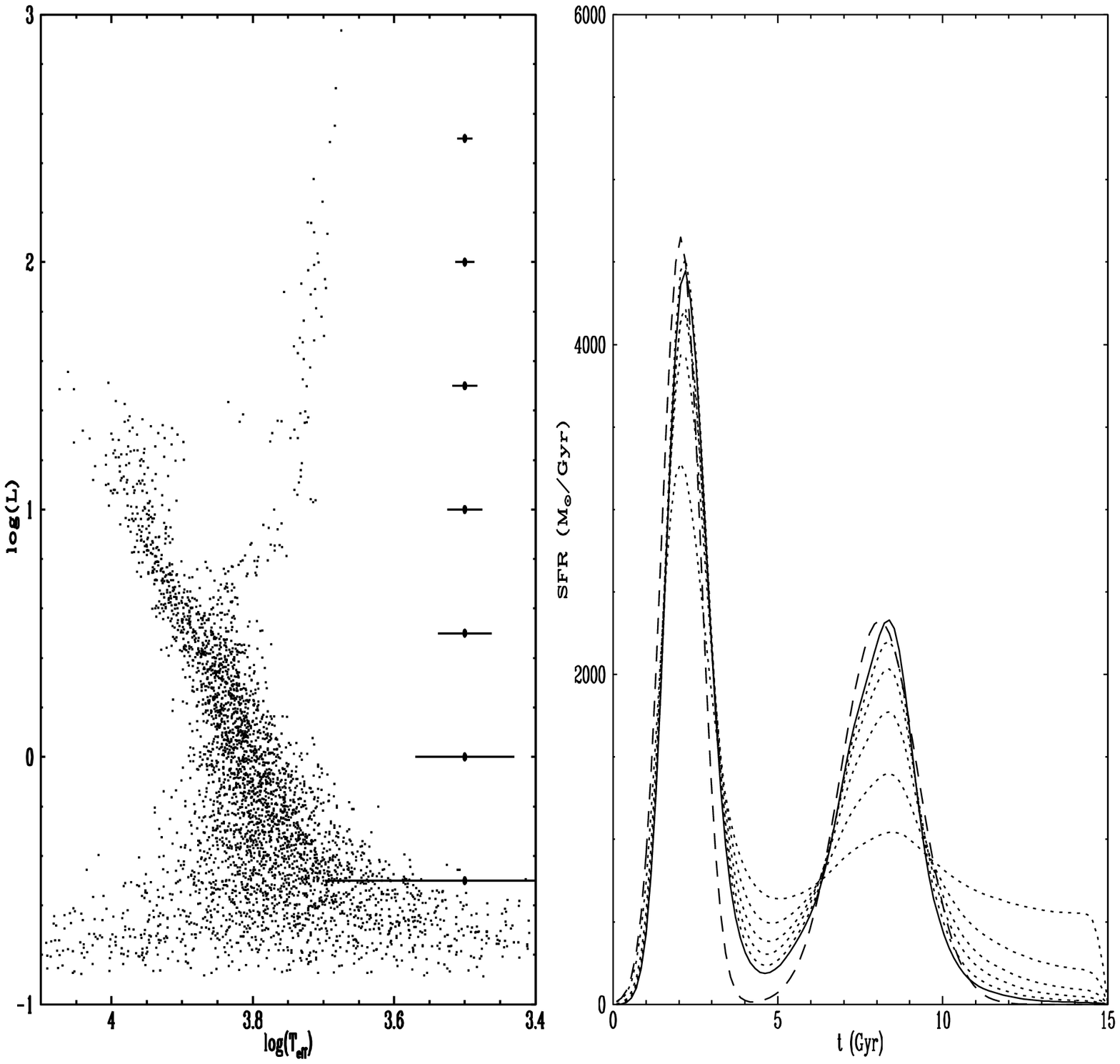,angle=0,width=18.1cm,height=9.0cm}
\@ \textbf{Figure 1.}\hspace{5pt}{
\begin{flushleft}\textbf{Left:} Synthetic HR diagram 
resulting from the first input $SFR(t)$. The length of the bars to the right
corresponds to $2\sigma$ in the $log(T_{eff})$ error to each side of the dots.
\textbf{Right:} First input $SFR(t)$, dashed line. Also shown
are the derived $SFR(t)$'s after 2, 4, 6, 8 and 10 iterations of the inversion method, 
dotted curves. The 12th iteration is given by the solid curve, showing 
convergence and a reliable recovery of the input $SFR(t)$. 
\end{flushleft}}
\end{figure*} 

In order to see whether the iterative method outlined in the appendix
converges, and if so, whether it does so to the correct $SFR(t)$, we
shall perform a variety of tests using synthetic HR diagrams produced
from known $SFR(t)$'s . To produce a realistic HR diagram from a
proposed $SFR(t)$ requires firstly a method of obtaining the colour
and luminosity of a star of a given mass and age. Interpolating
between isochrones is a risky procedure which can imprint spurious
structure in the inference procedure, given the almost discontinuous
way that stars' properties vary across critical points along the
isochrones, and how these critical points vary with time and metallicity.  
To avoid this we use the latest Padova (Fagotto et al. 1994, Girardi et al. 1996) 
full stellar tracks, calculated at
fine variable time intervals, and a careful  interpolating method which uses
only stars at constant evolutionary phases to construct an isochrone
library. We calculate 100 isochrones  containing 1000 uniformly spaced
masses each, with a linear spacing between 0.1 and 15 Gyr, which determines
the time resolution with which we implement the method to be 150 Myr. An
arbitrary time resolution can be achieved using a finer isochrone grid,
which only increases the calculation times linearly with the number of
intervals. Unless otherwise stated, we shall be assuming a metallicity
of $[Fe/H]_{\odot}=-1.7$, which is in the range of the spectroscopically
derived values for the dSph's (Grebel 1997). Although in comparison
with real data  it is important to use colours and magnitudes, trying
to make these first tests as clean as possible, we perform them on the
theoretical  HR diagram, in terms only of temperature (T) and
luminosity (L). Units throughout are $L_{\odot}$, degrees $K, t(Gyr)$ and 
$M_{\odot}/Gyr$.

Having fixed the isochrones, we now need to specify the manner in which
the density of stars will vary along these isochrones, i.e. an IMF. 
We use the IMF derived by Kroupa et al. (1993), where a single fit to
this function is seen to hold for stars towards both Galactic poles,
and for all stars in the solar neighborhood. In analyzing the stellar 
distribution towards the Galactic poles, a wide range of metallicities
and ages is sampled, and care was taken to account for all the
effects this introduces, including the changing mass-luminosity
relation at different ages and metallicities, completeness effects as a
function of luminosity and distance, and the contribution of binaries.
At this point we shall assume their result to be of universal
validity, and use their fit:

\begin{equation}
\rho(m) \propto \left\{
\begin{array}{rl}
     m^{-1.3} &  0.08M_{\odot} <m\le 0.5 M_{\odot} \\[1.0 ex] 
     m^{-2.2} &  0.5M_{\odot} <m\le 1.0 M_{\odot} \\[1.0 ex]
     m^{-2.7} &  1.0M_{\odot} < m
\end{array}\right.
\end{equation} 

We normalize this relation such that a unit total mass is contained
upwards of $0.08 M_{\odot}$, although only stars in the mass range
$0.6 - 3 M_{\odot}$ can end up in the HR diagram.
We can now choose a $SFR(t)$, and use the IMF of equation~(9) to populate our
isochrones and create a synthetic HR diagram, after including
``observational'' errors, assumed as Gaussian on log(T).
The dispersion is assumed to depend only on $L$, and as an illustrative
example we will use:

\begin{equation}\label{Sigma}
\sigma(L) ={0.035\over \left[ log(L)+1 \right]^{1.5} }
\end{equation} 

\subsection{A simple 2-burst example}

As a first test
we use a $SFR(t)$ consisting of two Gaussian bursts at different
epochs, of different amplitudes and total masses. This $SFR(t)$ is 
shown by the dashed line in the right panel of Figure(1),
where the time axis shows the age of the corresponding stellar populations. 
The left panel of Figure~(1) shows the resulting HR diagram which contains a total of 3819 stars. 
To ensure a realistic error structure 
the shape of equation~(\ref{Sigma}) was obtained from a fit to the errors of the HST 
observations of dSph galaxies of Unavane and Gilmore (private communication). The amplitude of
this error is representative of what is seen in current HST observations.
From the synthetic HR diagram the general
features of the input $SFR(t)$ can be seen, in that two basic
populations are evident. Obtaining the precise duration and location of
these two bursts requires more work, and the detailed shape of each 
is quite hard to recover.

From the position of every one of the 1324 simulated stars 
with $log(L)>0$ on Figure~(1) (see below) we construct
the matrix $G_{i}(t)$, where we further assume that the
``observational'' errors are well understood i.e. $\sigma(L)$ is
known. Since the colour of a star having a given luminosity can
sometimes be a multi-valued function,  in practice we check along 
a given isochrone, to find all possible masses a given observed star
might have as a function of time, and add all contributions (mostly 1,
sometimes 2 and occasionally 3) in the same  $G_{i}(t)$. Calculating
this matrix is the only  slow part of the procedure, and is equivalent
to calculating the likelihood of one model. The likelihood matrix $G_{i}(t)$
is the only input required by the method. The total number of stars
is used as a normalization constraint at each iteration, needed to
recover $SFR(t)$ from $Y(t)$. As mentioned earlier, 
it is not necessary to calculate the likelihood over the solution
space being considered, i.e. $G_{i}(t)$ is only calculated once, which
makes the method highly efficient. 

Given the degeneracy of isocrones of different ages in the main sequence region,
the lower fraction of the HR diagram is of relevance only in establishing the
total normalization condition, and not in determining the shape of the $SFR(t)$.
For this reason, we only include in the inference procedure stars with $log(L)>0$,
other stars are only used in fixing the overall normalization. The final results are not
affected by this cut, but the iterative procedure converges much more rapidly and in 
a numerically more stable way if the lower degenerate and high error region of the HR
diagram is excluded.
	
\begin{figure*}
\epsfig{file=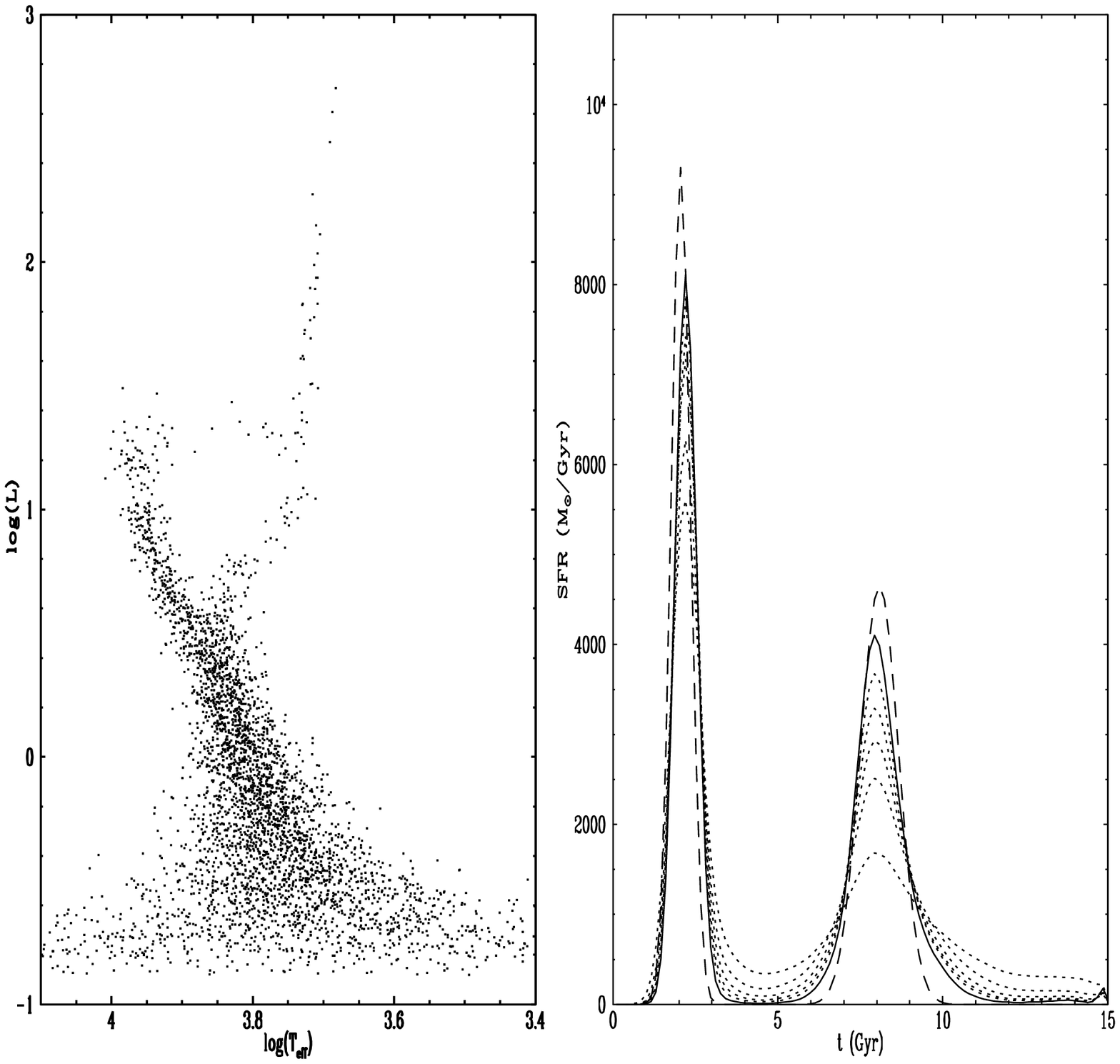,angle=0,width=18.1cm,height=9.0cm}
\@ \textbf{Figure 2.}\hspace{5pt}{
\begin{flushleft}\textbf{Left:} Synthetic HR diagram 
resulting from the second input $SFR(t)$.
\textbf{Right:} Second input $SFR(t)$, dashed line. Also shown
are the  3, 6, 9, 12 and 15 iterations of the inversion method, dotted curves.
The 20th iteration is given by the solid curve, showing 
convergence and a reliable recovery of the input $SFR(t)$.
\end{flushleft}}
\end{figure*}

In Figure~(1) we also show the results of the first 12 iterations
of the method every 2 iterations, which form a sequence of increasing
resemblance to the input $SFR(t)$. The distance between successive
iterations decreases monotonically at all ages, which together with
the fact that after 12 iterations no further change is seen, shows the
convergence of the method for this case. From the 2nd iteration (lowest dotted curve
in the burst regions) it
can be seen that the iteration of the variational calculus equation
constructed from maximizing the likelihood is able to recover the
input $SFR(t)$ efficiently. The positions, shapes and relative masses
of the two bursts were correctly inferred by the 2nd iteration,
although it took longer for the method to eliminate the populations
outside of the two input bursts. The convergence solution is in
remarkable agreement with the input $SFR(t)$, and only differs
slightly, as seen from Figure~(1). No information
was used in the inverting procedure beyond that which is available from
the synthetic HR diagram, which was used extensively in constructing the likelihood
matrix $G_{i}(t)$, which is the only input required by the inversion. The variational
calculus method recovers a $SFR(t)$ for which the first variation of
the likelihood vanishes, without assuming any {\it a priori} condition
on the $SFR(t)$, beyond being a continuous twice differentiable function of time.

\subsection{Testing temporal resolution}

The second test uses an input $SFR(t)$ which differs from the previous one
in that the bursts are of much shorter duration and larger amplitude, to 
approximately preserve the total number of stars. The input $SFR(t)$ of
this case is shown by the dashed curve in the right panel of Figure~(2).
The HR diagram which results from this $SFR(t)$ is shown in the left panel
of Figure~(2) and shows basically the same populations as in Figure~(1)
but with a much smaller spread, it contains a total of 3783 stars, with 1299
above $log(L)=0$. The errors in these two cases were equal.

\begin{figure*}
\epsfig{file=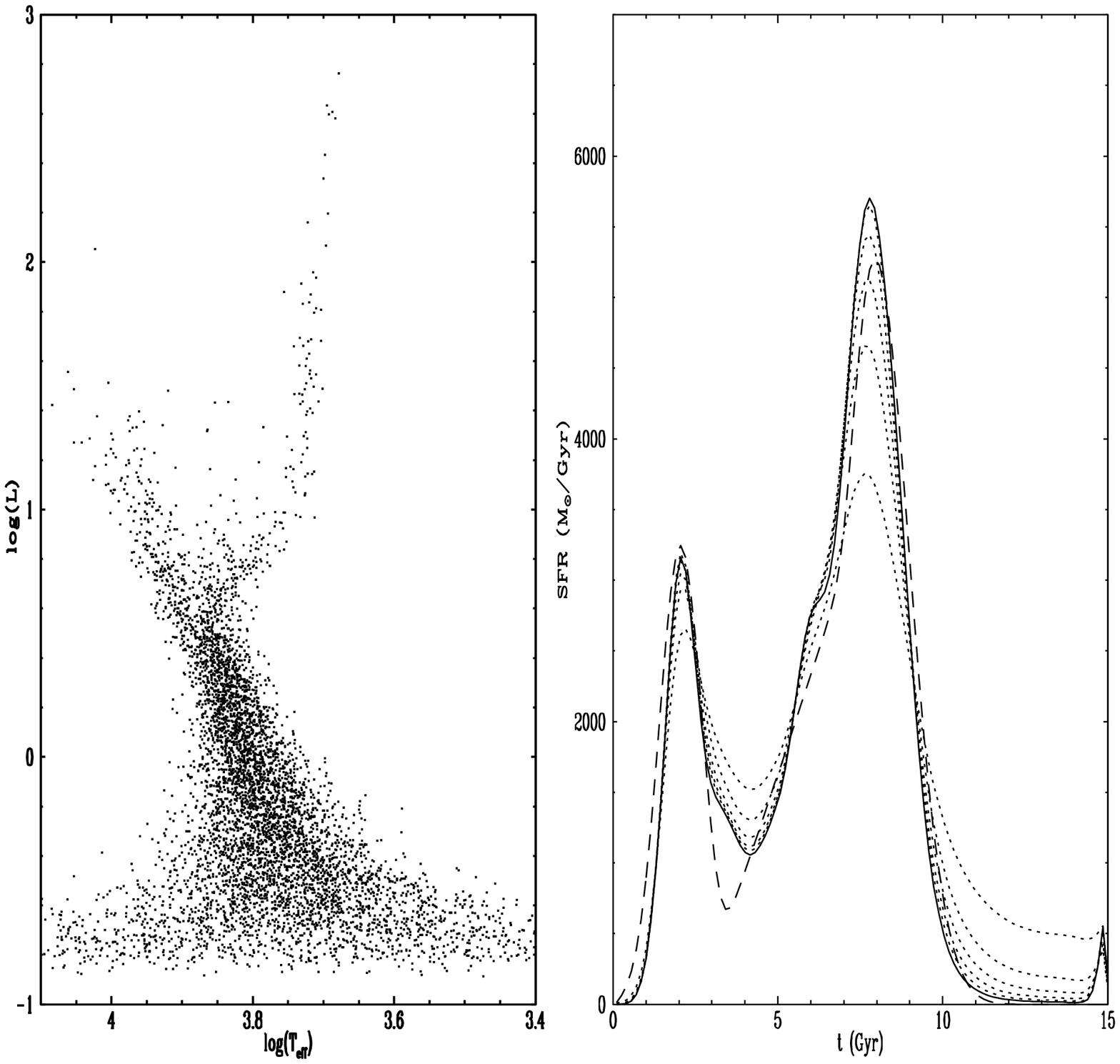,angle=0,width=18.1cm,height=9.0cm}
\@ \textbf{Figure 3.}\hspace{5pt}{
\begin{flushleft}\textbf{Left:} Synthetic HR diagram 
resulting from the third input $SFR(t)$.
\textbf{Right:} Third input $SFR(t)$, dashed line. Also shown
are the  3, 6, 9, 12 and 15 iterations of the inversion method, dotted curves.
The 18th iteration is given by the solid curve, showing 
convergence and a good recovery of the input $SFR(t)$.
\end{flushleft}}
\end{figure*}

The inversion procedure is shown in the right panel of Figure~(2), were it
can be seen that the convergence of the method remains robust, although this
time it took slightly longer, the first 15 iterations are shown every 3. The
$SFR(t)$ to which the method converged (after 20 iterations) again accurately reproduces the input one,
the age, duration, amplitude and shape of the two input bursts were correctly
inferred. That the shorter duration of the bursts was correctly inferred in this case
shows that in the previous one the reconstructed age duration was not due to the spread
caused by the errors, but was actually resolved in the data, and recovered 
correctly by the method. In this second case however, the spread due to the errors begins to be
comparable to the intrinsic one of the input $SFR(t)$, and causes an artificial
broadening of the recovered stellar ages, particularly in the older component.
This last effect causes also a slight underestimate in the maximum amplitude of the bursts.
Reducing the duration of the older component further would not produce a shorter duration
in the inferred burst, unless the errors were also reduced. That is, the method is capable
of recovering the full age precision allowed by the observational errors.

\subsection{Burst plus continuum cases}

As a third test we present Figure~(3), which is analogous
to Figures~(1) and (2). The synthetic HR diagram this time contains
5907 stars with 1913 upwards of $log(L)=0$, and has the same errors as were used in Figure~(1). From
the HR diagram in Figure~(3) it can be seen that we have a dominant old population and
a more recent component, with some degree of intermediary activity. In
order to
go beyond such a qualitative description to a quantitative assessment
of detailed shapes and magnitudes of the star formation episodes we
apply our inversion method, with results shown in the right panel of Figure~(3), this time
the first 18 iterations, every 3. 

The dashed line in Figure (3) shows the input $SFR(t)$, which
consists of a young narrow burst, a linearly decreasing intermediary component, and
a larger older one. As Figure (3) shows, the method successfully
inverted the HR diagram, converging very rapidly into the main
features of the input $SFR(t)$. The youngest star formation episode
was recovered remarkably well, as was the older dominant component, the
$SFR(t)$ to which the method converges differs only marginally from the
input one in these two regions. In the intermediary region the method
did recover the main features of the input $SFR(t)$, although the minimum
was slightly overestimated. A low level numerical error due to edge effects
near $t=15 Gyr$ also appeared.
A general feature of the method is that it rapidly
eliminates any population younger than those present, but it takes
longer to exclude populations older than those present. In general,
the resolution of the inversion procedure decreases with age, as the
critical regions of older isochrones become swamped in the noise. Lower
noise levels make the inversion procedure easier and more accurate
results are obtained. 

As a fourth test we use an input $SFR(t)$ which differs from that of
the previous test, only in that the intermediary
population increases with time, rather than the opposite. The
resulting HR diagram is therefore very similar to the previous case, as can be seen from
comparing the left panels of Figures (3) and (4). The errors were the same as in
all previous cases, while this test produced 5235 stars, 1743 in the region used by the
inference procedure. 
Figure (4) shows the input $SFR(t)$
as the dashed line, and the first 15 iterations of the inversion
method, every 3, together with the final result at 20 iterations. 
As the ages and amplitudes of the two main components match those
of the previous case, the HR diagrams of Figures (3) and (4) differ only by a few 
stars. That the method is capable of discriminating between
two such similar cases is encouraging. From the solid curve in Figure (4) it can be
seen that again the method converges, and again into a $SFR(t)$ which
reproduces the input one to a high degree, the method
efficiently eliminated populations outside of those actually present.
Not only are the main components identified in age and amplitude, but also the
intrinsic shape of the bursts is recovered, as in the abrupt end of the younger 
burst, and the slanting intermediary component. Moderate
discrepancies between the input and recovered $SFR(t)$ remained,
specially in the appearance of an artificial oscillatory
component superimposed mostly on the slanting component, possibly
due to structure introduced by the small number of stars driving the likelihood matrix in that region.

\begin{figure*}
\epsfig{file=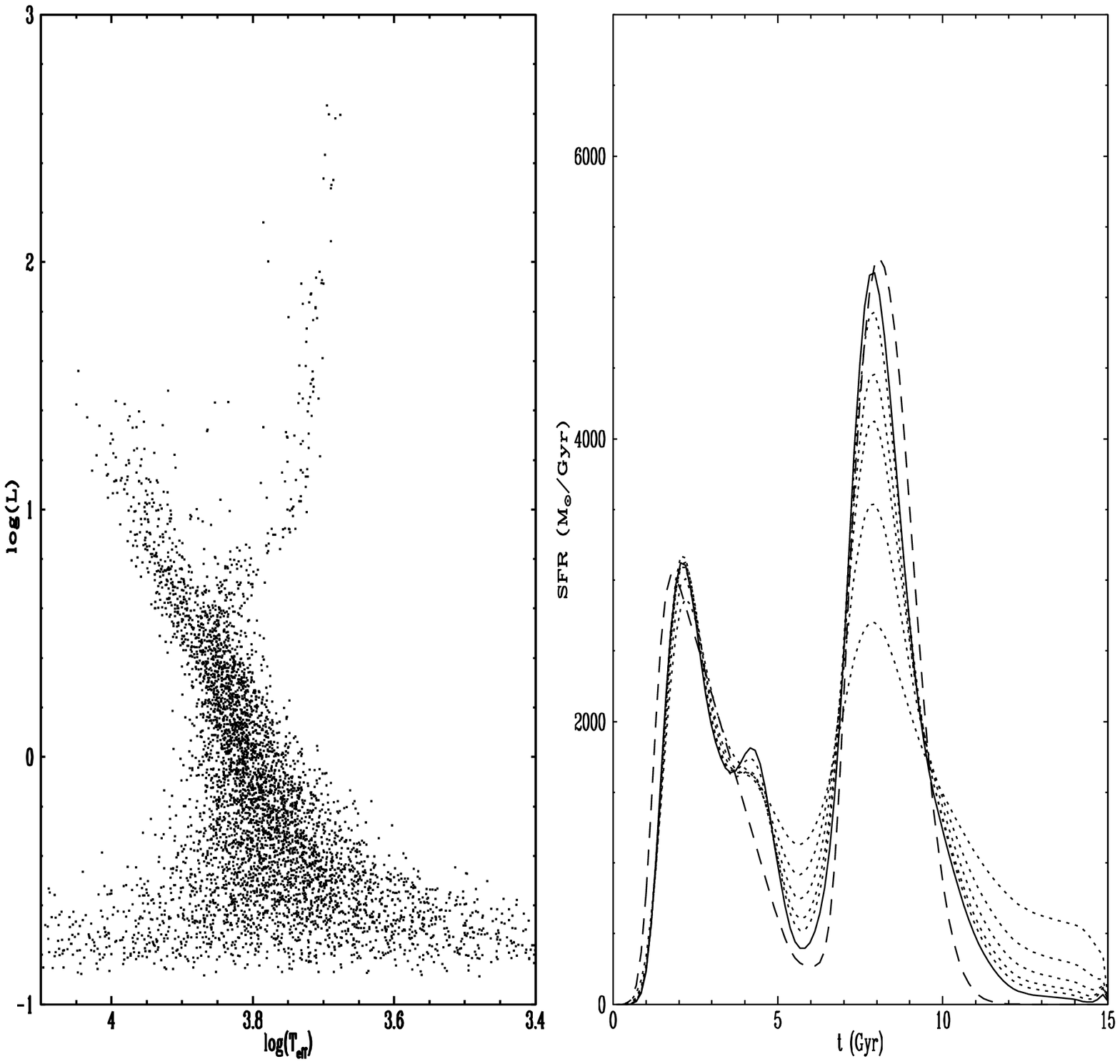,angle=0,width=18.1cm,height=9.0cm}
\@ \textbf{Figure 4.}\hspace{5pt}{
\begin{flushleft}\textbf{Left:} Synthetic HR diagram 
resulting from the fourth input $SFR(t)$.
\textbf{Right:} Fourth input $SFR(t)$, dashed line. Also shown
are the  3, 6, 9, 12 and 15 iterations of the inversion method, dotted curves.
The 20th iteration is given by the solid curve, showing 
convergence and a good recovery of the input $SFR(t)$. 
\end{flushleft}}
\end{figure*}

\begin{figure*}
\epsfig{file=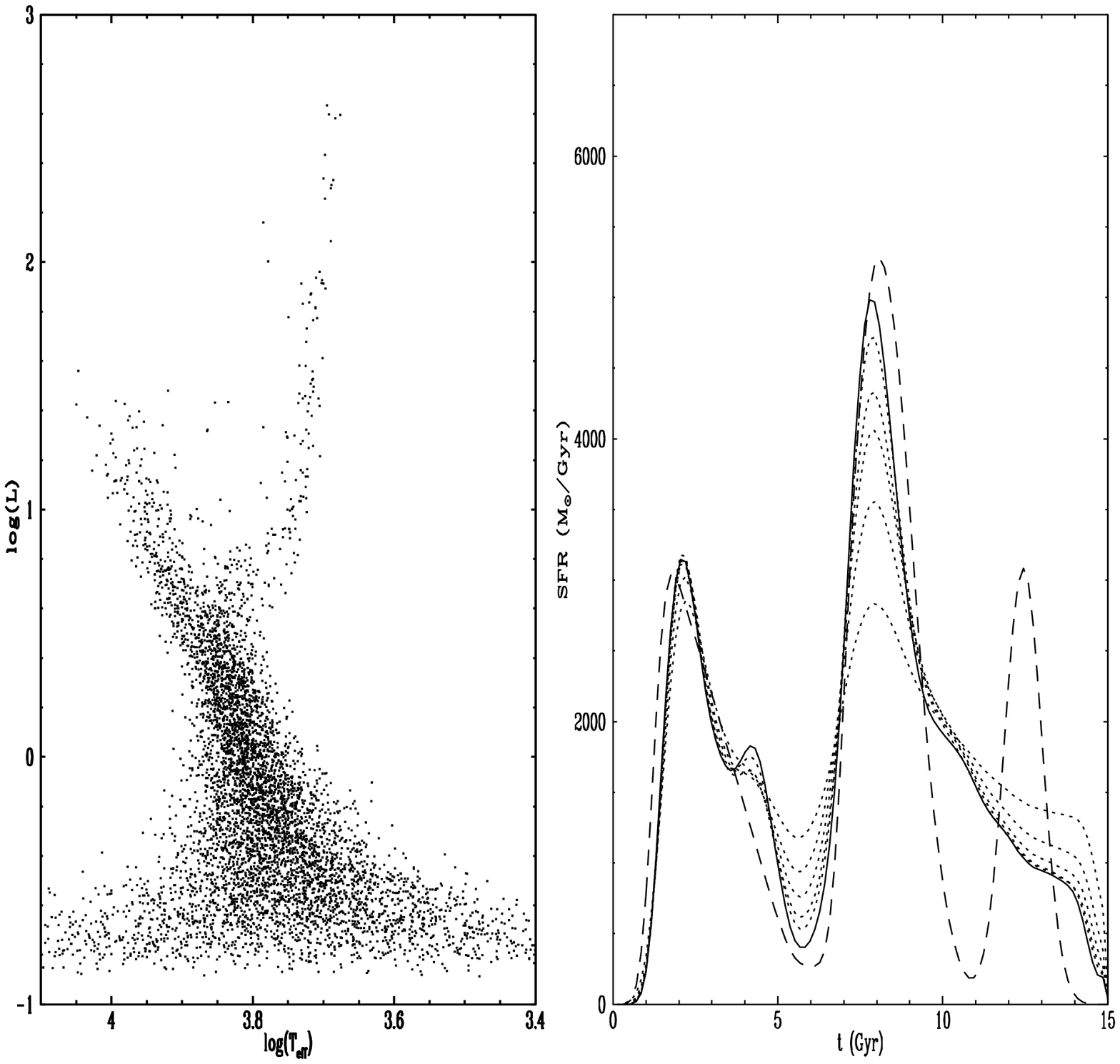,angle=0,width=18.1cm,height=9.0cm}
\@ \textbf{Figure 5.}\hspace{5pt}{
\begin{flushleft} \textbf{Left:} Synthetic HR diagram 
resulting from the fifth input $SFR(t)$.
\textbf{Right:} Fifth input $SFR(t)$, dashed line. Also shown
are the  3, 6, 9, 12 and 15 iterations of the inversion method, dotted curves.
The 20th iteration is given by the solid curve, showing 
convergence and a good recovery of the input $SFR(t)$ for $t< 10 Gyr$.
\end{flushleft}}
\end{figure*}

\begin{figure*}
\epsfig{file=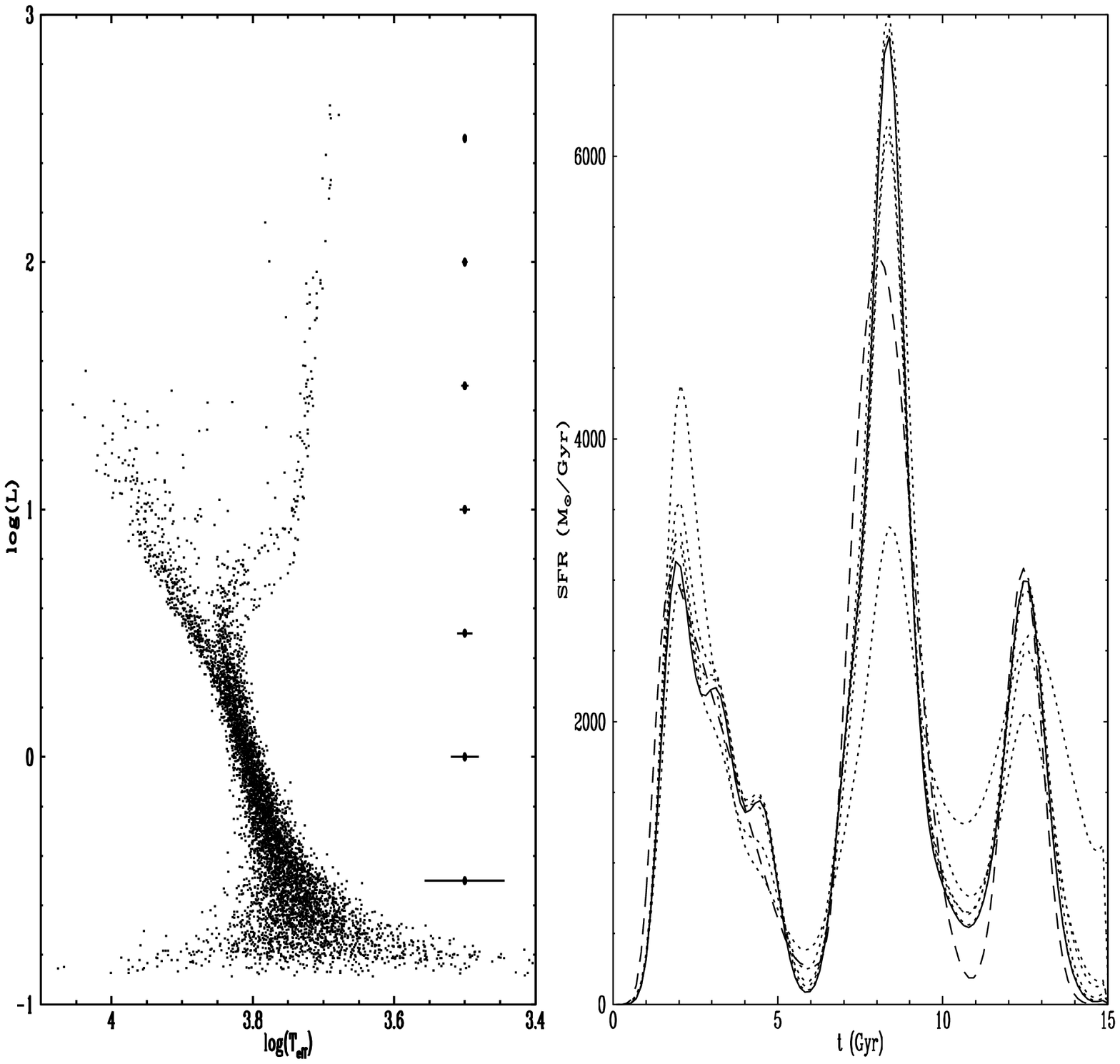,angle=0,width=18.1cm,height=9.0cm}
\@ \textbf{Figure 6.}\hspace{5pt}{
\begin{flushleft}\textbf{Left:} Synthetic HR diagram 
resulting from the fifth input $SFR(t)$, produced using a much lower noise level.
As in Figure~(1), the length of the bars to the right corresponds to $2\sigma$
in the $log(L)$ error to each side of the dots.
\textbf{Right:} Fifth input $SFR(t)$, dashed line. Also shown
are the  2, 4, 6, 8 and 9 iterations of the inversion method, dotted curves.
The 10th iteration is given by the solid curve, showing rapid 
convergence and a good recovery of the input $SFR(t)$.
\end{flushleft}}
\end{figure*}

\subsection{Very old populations}

The fifth test explores explicitly the way in which the method reacts to
populations older than 10Gyr, an approximate limit beyond which 
observational errors totally confuse the turn off points, in
situations equivalent to current HST observations of dSph galaxies.
This fifth test has the same input $SFR(t)$ as shown in Figure (4),
with the difference that an old component was added beyond 10
Gyr. This $SFR(t)$ is shown by the dashed line in the right panel of Figure (5). 
The left panel of Figure (5) shows the resulting HR diagram 
which is used in the inversion process, and which only marginally differs from the one of Figure (4).

Figure (5) shows that the inversion procedure converged to a $SFR(t)$
which is a highly accurate representation of the input $SFR(t)$ in
regions younger than around 10 Gyr. Although the result of this test
differed from that of the previous one for populations older than 10 Gyr in
that the stars from this region were detected, the time structure of the oldest burst was totally
erased. The poorer performance of the method in this region
is also due to the proportionally smaller numbers of stars which live
into the $t=0$ HR diagram. If more stars had been available in this
last test, the older regions would have been better reproduced. 

\subsection{Sensitivity to noise}

In this section we present Figure (6), which has the same
input $SFR(t)$ as in case 5, and differs only in that a much lower noise
level was assumed. In constructing the HR diagram seen in the left panel
of Figure (6) the numerical constant in eq (10) was reduced from $0.035$ to
$0.01$. This lower noise level is reflected in the clearer HR diagram, where the
older population is now distinguishable from the noise of the younger main sequence. 
The right panel in Figure (6) shows the result of the inversion procedure, which differs from case 5
mostly in the speed with which the method converged, only 10 iterations were needed. 
The few stars in the oldest component which can be separated from the younger main
sequence are sufficient to accurately recover the shape for this burst.

In general, the variational calculus treatment of the maximum likelihood
problem, together with the iterative method for solving the resulting
equation, works remarkably well, within the practical limits set by the
``observational'' errors. Having assumed only that the $SFR(t)$ was a
continuous function of time, the method manages to recover the input
function quite accurately, under conditions similar to those present
in current observations. In forthcoming papers we apply this methodology to derive 
star formation histories for several Galactic satellites, and the Solar Neighbourhood.

\subsection{Sensitivity to initialization procedure}

\begin{figure*}
\epsfig{file=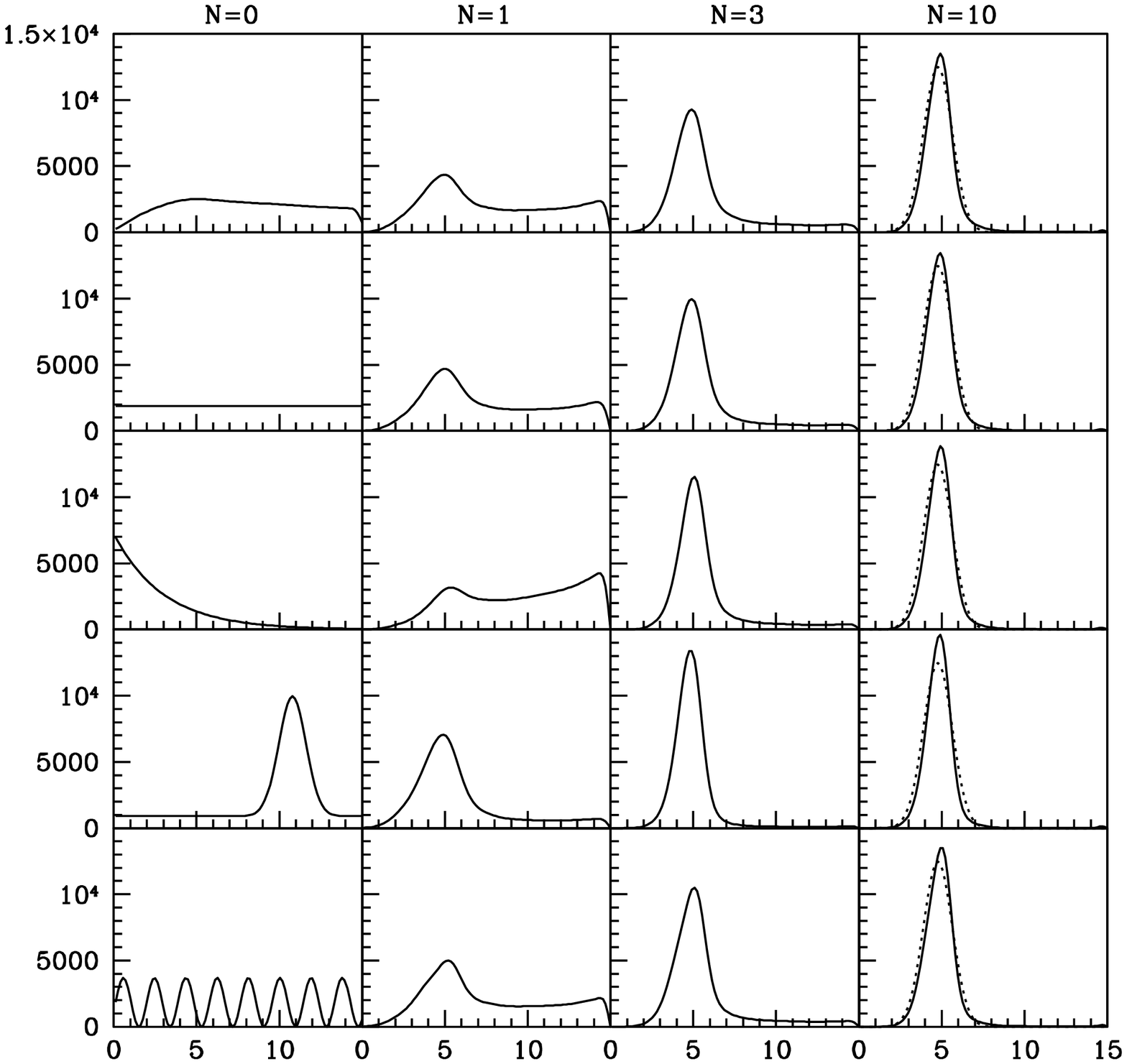,angle=0,width=18.1cm,height=18.1cm}
\@ \textbf{Figure 7.}\hspace{5pt}{
\begin{flushleft} The first column in this figure shows the initial $SFR(t)'s$ from which the method
started, in the inversion of the HR diagram resulting from the $SFR(t)$ shown in the last column by the
dotted line. The following columns show the first, third and tenth iterations of the method, showing the result
to be totally independent of the initialization procedure, as well as being highly accurate. 
The top row shows the ``natural'' initialization for the problem, see text.

\end{flushleft}}
\end{figure*}

\begin{figure*}
\epsfig{file=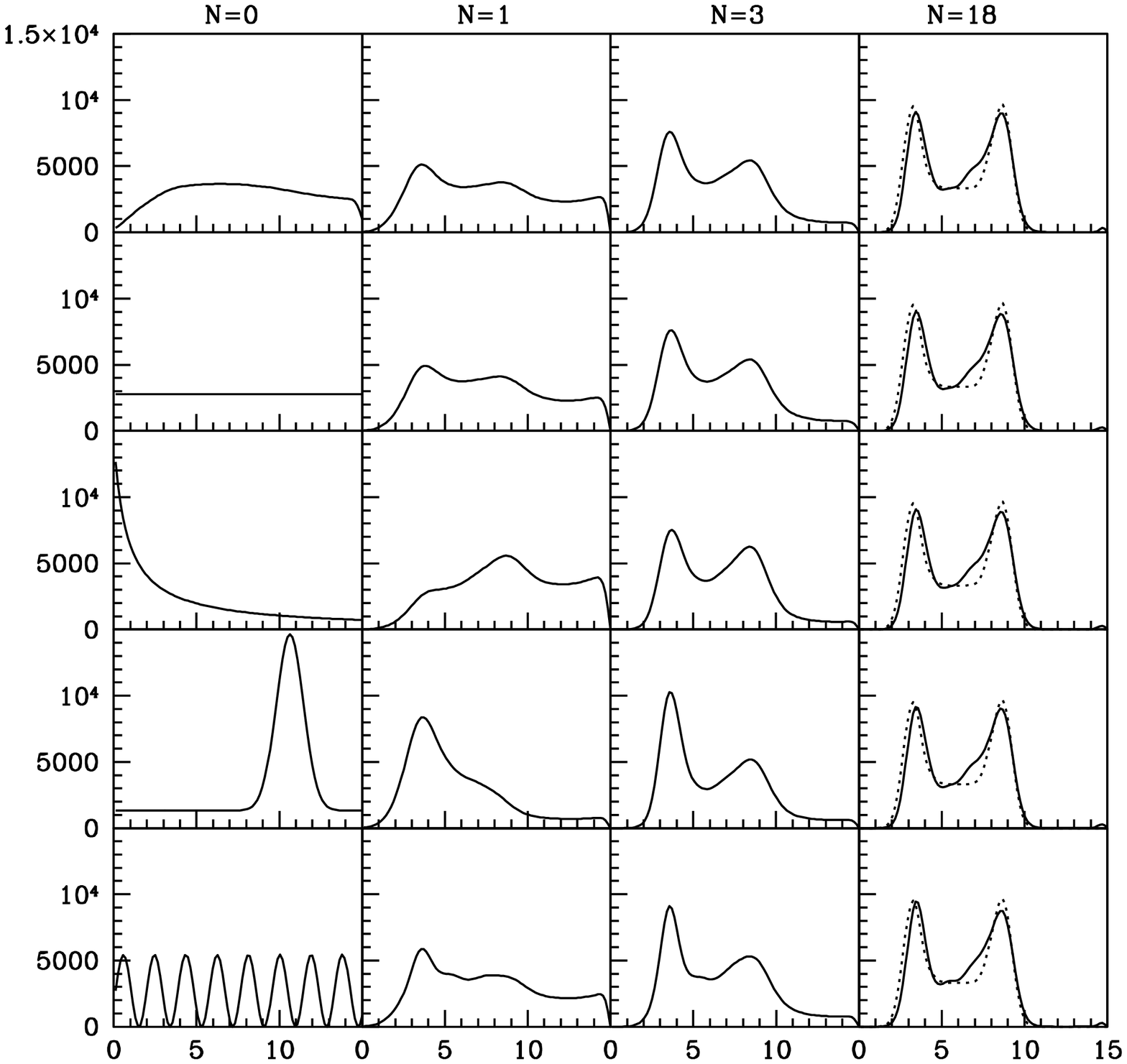,angle=0,width=18.1cm,height=18.1cm}
\@ \textbf{Figure 8.}\hspace{5pt}{
\begin{flushleft}The first column in this figure shows the initial $SFR(t)'s$ from which the method
started, in the inversion of the HR diagram resulting from the $SFR(t)$ shown in the last column by the
dotted line. The following columns show the first, third and eighteenth iterations of the method, showing the result
to be totally independent of the initialization procedure, as well as being highly accurate.
The top row shows the ``natural'' initialization for the problem, see text.

\end{flushleft}}
\end{figure*}

\begin{figure*}
\epsfig{file=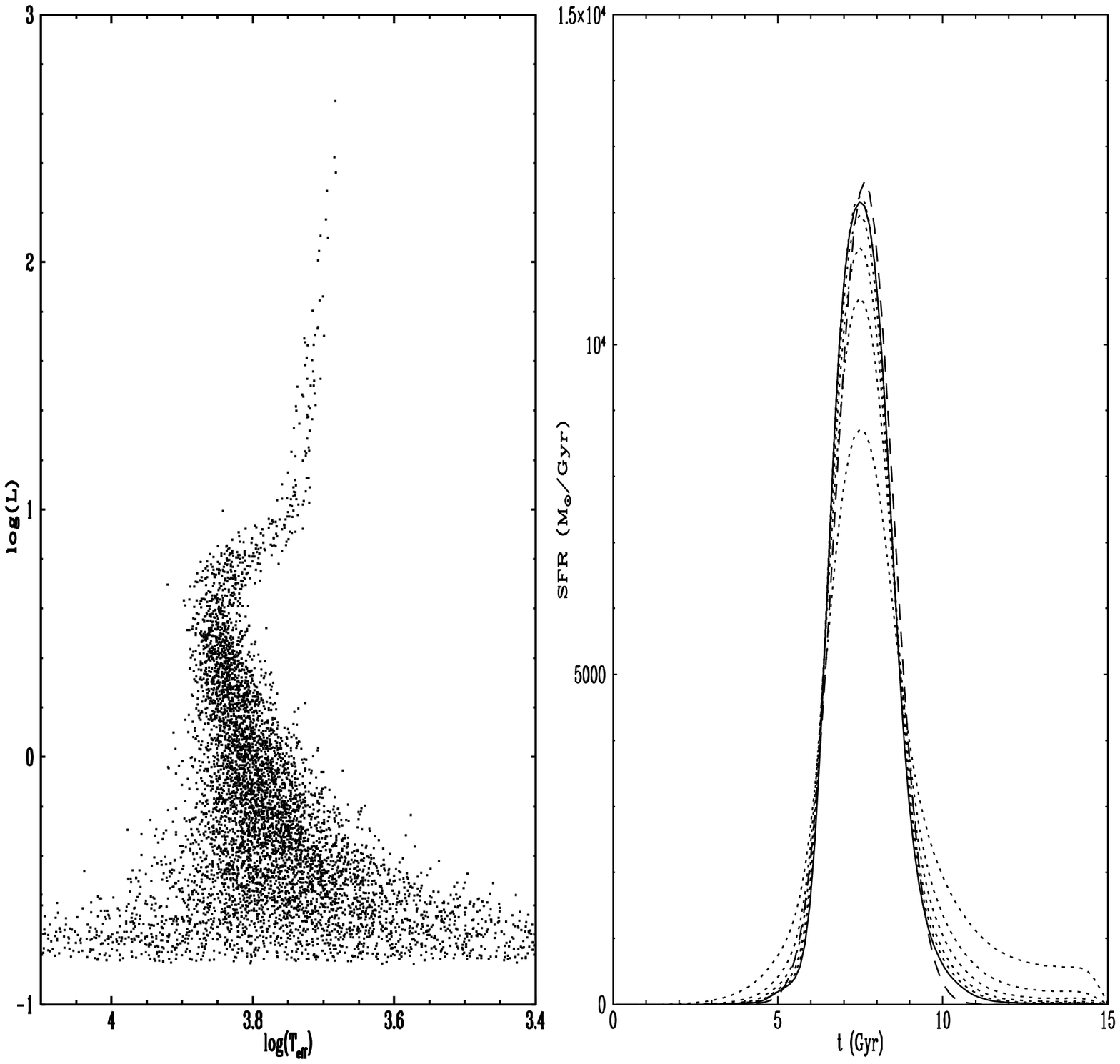,angle=0,width=18.1cm,height=9.0cm}
\@ \textbf{Figure 9.}\hspace{5pt}{
\begin{flushleft}\textbf{Left:}  Synthetic HR diagram 
resulting from the control input $SFR(t)$.
\textbf{Right:} Sixth input $SFR(t)$, dashed line. Also shown
are the  3, 6, 9, 12 and 15 iterations of the inversion method, dotted curves.
The 18th iteration is given by the solid curve, showing 
convergence and an accurate recovery of the input $SFR(t)$. This is used as a
control case against which to compare variations in the assumed
IMF, metallicity and binary fraction. 
\end{flushleft}}
\end{figure*}

As explained at the end of section 2, the integro-differential equation (equation 8) which
results from applying variational calculus to maximize the functional of
$SFR(t)$ given by the likelihood function (equation 6) is solved
iteratively. This requires the use of an initialization function for the method.
In all previous examples we have used the method described in the appendix, where
for the first iteration all the $I(i)$ coefficients are set equal to 1, and equation (8)
then solved using a finite differences procedure to obtain $SFR_{0}(t)$. This provides
a natural initialization for the problem which is driven by the likelihood matrix $G_{i}(t)$.
This same method will also be used in the following sections. 
In this section however, we study the generality of this initialization and the
effects of adopting different ones.
 
We present two cases, in each a fixed HR diagram was inverted using a variety
of initialization conditions, the first being the natural condition. The other
initialization procedures used where actually to start by supplying $SFR_{0}(t)$
directly, the rest of the procedure being unchanged. One should expect this
differences to matter little, as in the solving of equation (8) the
previous solution $SFR_{j}(t)$ enters in the determination of $SFR_{j+1}(t)$ only
through the $I_{j+1}(i)$ coefficients, i.e. only through time integrals
of $SFR_{j}(t)$, weighted by the likelihood matrix. In this way the details
of the previous iteration are lost to the next, it is the likelihood matrix and
the time derivative of it which drive the convergence (see equation 8).

Figure (7) shows the effects of adopting different initialization procedures
to invert the HR diagram resulting from a simple Gaussian burst input $SFR(t)$.
This input $SFR(t)$ resulted in 7411 stars, 2458 brighter than the $log(L)=0$ limit, the
noise level was the same as in Figure (1).
In the first row the natural condition (N) was used, where $SFR_{0}(t)$, $SFR_{1}(t)$,
$SFR_{3}(t)$ and $SFR_{10}(t)$ are shown in the four panels of the first row.
The last panel also shows the input $SFR(t)$ as a dotted curve. In the second
row a constant $SFR_{0}(t)$ was used (C), a decaying one in the third row (D), a
misplaced burst in the fourth (M) and a sine wave (S) in the fifth row, all sequences
show the initial $SFR(t)$  and the first, third and tenth iterations of equation (8).
In this way the first column shows the initial, the second the first iteration, the third the
third iteration, and the last column the tenth iteration, by which time the method 
had converged in all cases.
Clearly, as each iteration enters into the calculation of the next only through an integral
condition, the method is highly independent of the choice of initialization function. Although
the five initial $SFR_{0}(t)$ vary markedly, already the first iteration shows a much 
more homogeneous sample, whilst by the third iteration all initialization procedures
give highly similar answers. As expected, the final result is seen to be totally independent
of the initialization procedure, it is driven exclusively by the likelihood matrix. This
last point is shown in the last column of Figure (7), where the final answers of the five
different initial $SFR_{0}(t)$ can be seen to differ only marginally, and to be in all cases a 
very good estimate of the input $SFR(t)$, which is shown by the dotted curve.

Figure (8) is analogous to Figure (7), but was constructed using a more complicated input
$SFR(t)$, shown by the dashed curve in the last column of Figure (8). Again, even though
the initial $SFR_{0}(t)'s$ where very different, the second and third columns show the variation 
between the first iterations to be much reduced, and almost gone by the third. The final result
in all cases is identical, and a good approximation to the input $SFR(t)$, which was more complex
than what was used in Figure (7). Even the slight error seen between the final result and the input 
$SFR(t)$ is a feature independent of the initialization assumptions. The solution 
is clearly seen to be driven by the likelihood 
matrix, which is of course the same for all cases, as it depends only on the position of the stars in the 
HR diagram, and the stellar evolutionary model used. 
Case (S) illustrates how spurious high frequency structure 
does not propagate between iterations, which is natural as only integral coefficients
of $SFR_{j}(t)$ intervene in estimating $SFR_{j+1}(t)$. This is one of the merits of
our method, as numerical noise is suppressed. The independence of the answer on the 
initialization procedure is encouraging of the solid statistical foundation of the
method and the validity of the variational calculus approach, but is not an essential
feature, as there exists a natural initialization internal to the problem, which should
be preferred in all cases. Henceforth we shall make use only of this natural initialization
as described in the appendix.

\section{Sensitivity to uncertainties in IMF, metallicity and binaries}

\begin{figure*}
\epsfig{file=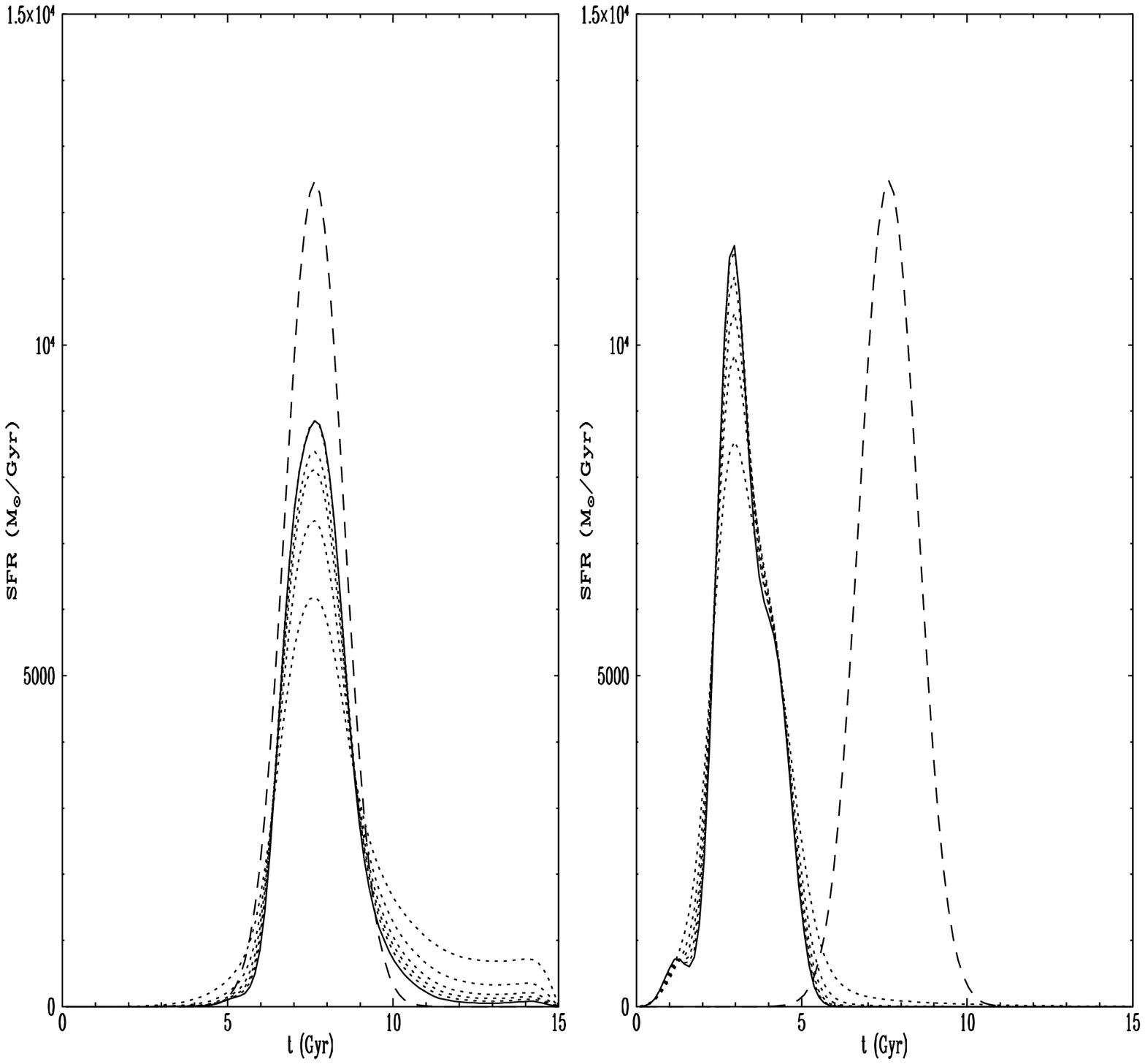,angle=0,width=18.1cm,height=9.0cm}
\@ \textbf{Figure 10.}\hspace{5pt}{
\begin{flushleft}\textbf{Left:} Sixth input $SFR(t)$, dashed line. Also shown
are the  3, 6, 9, 12 and 15 iterations of the inversion method, dotted curves.
The 18 iteration is given by the solid curve, assuming an IMF
much more weighted towards smaller masses than the one used for the HR diagram, which
produces a normalization error. \textbf{Right:} Sixth input $SFR(t)$, dashed line. Also shown
are the  3, 6, 9, 12 and 15 iterations of the inversion method, dotted curves. 
The 18th iteration is given by the solid curve, assuming a
metallicity one order of magnitude higher than the one used for the HR diagram,
which results in the method converging to a younger $SFR(t)$. 
\end{flushleft}}
\end{figure*}

Uncertainties in the IMF, metallicity and binaries
differ from simple sample size or photometric error in inducing a 
systematic mismatch between the isochrones used in any specific calculation
and those which describe the astrophysics of the HR diagram being inverted.
The following tests show the sensitivity of
the method to uncertainties in the input IMF, metallicities and 
binary fraction. Firstly
we present Figure (9), the synthetic HR diagram contains 6340 stars with 1808
brighter than $log(L)=0$, and
the input $SFR(t)$ is shown by the dashed line in the right panel of Figure (9). The
inversion of this HR diagram clearly shows again the
convergence of the method to an accurate representation of the input $SFR(t)$. 
This test was included to define a control case to which variations
can be compared.

\subsection{IMF mismatch}

The left panel in Figure (10) shows the results of applying the
inversion procedure with a ``wrong'' IMF to the HR diagram in
Figure (9) which was produced using the IMF of equation~(9). The inversion
procedure assumed an IMF $\rho(m) \propto m^{-3}$ for all masses. Comparing
Figure (9) and the left panel in Figure (10) it can be seen that the main effect of an error
in the IMF is a distorted normalization, which is based on
the total number of stars. The IMF used in the inversion is one in which low mass stars 
are more dominant than in the IMF used to construct the HR diagram, as a result, a lower
$SFR(t)$ sufficed to produce the correct number 
of stars. As the colour and luminosity of a star of a
given age and mass are not affected by changes in the IMF, the
location of the relevant populations was not 
affected. The net result of changing the IMF used in the inversion
was a reduction of the recovered $SFR(t)$ by a factor of $0.85$, which is 
the factor by which the mass in stars in the mass region which we are
sampling differs between the two IMF's. The convergence 
of the method was similarly unaffected. We
can conclude that errors in the IMF used, within the expected
uncertainties, do not affect the derived star formation history significantly.

\subsection{Metallicity mismatch}

In the following tests we investigate the effects of an uncertainty in
the metallicity; the well known degeneracy between the inferred
age and metallicity of an observed stellar population will be evident. Figure (10), 
right panel
shows the result of inverting the HR diagram of Figure (9), which was produced
using a metallicity of $[Fe/H]=-1.7$, using this time 
isochrones for $[Fe/H]=-0.7$ in the inversion procedure. The convergence 
of the method was not affected, and proceeded at the same rate as in the previous two cases,
 as the same HR diagram was used. The result of having
assumed a metallicity one order of magnitude higher than that of the stellar
population being inverted is a $SFR(t)$ much younger than the input one, 
as can be seen from Figure (10), right panel. This discrepancy is due to the fact that the
isochrones used in the inversion have very different temperatures and luminosities
for stars of a given mass, from those of the isochrones used to generate the HR diagram.
Actually, the colours and luminosities of stars from the higher metallicity
isochrones approximately correspond to those of younger stars from the lower metallicity
isochrones, the age-metallicity degeneracy. Having used isochrones in the inversion
procedure which do not correspond to the stars being studied also confuses the method
and the shape of the recovered $SFR(t)$ is slightly distorted. The younger age assigned
to the stars being analyzed also produces a slightly lower total $SFR(t)$, as with a
younger population a larger fraction of the stars live into the present day HR diagram.
Using real data, the slope of the RGB can be used to determine the metallicity
to greater accuracy than the difference between the used and assumed metallicities of this
example, where the effect was exaggerated.
It is interesting to perform the complementary test, where the HR diagram is produced
using a non constant metallicity, and inverted assuming a constant one. This is
presented in Figure (11).

\begin{figure*}
\epsfig{file=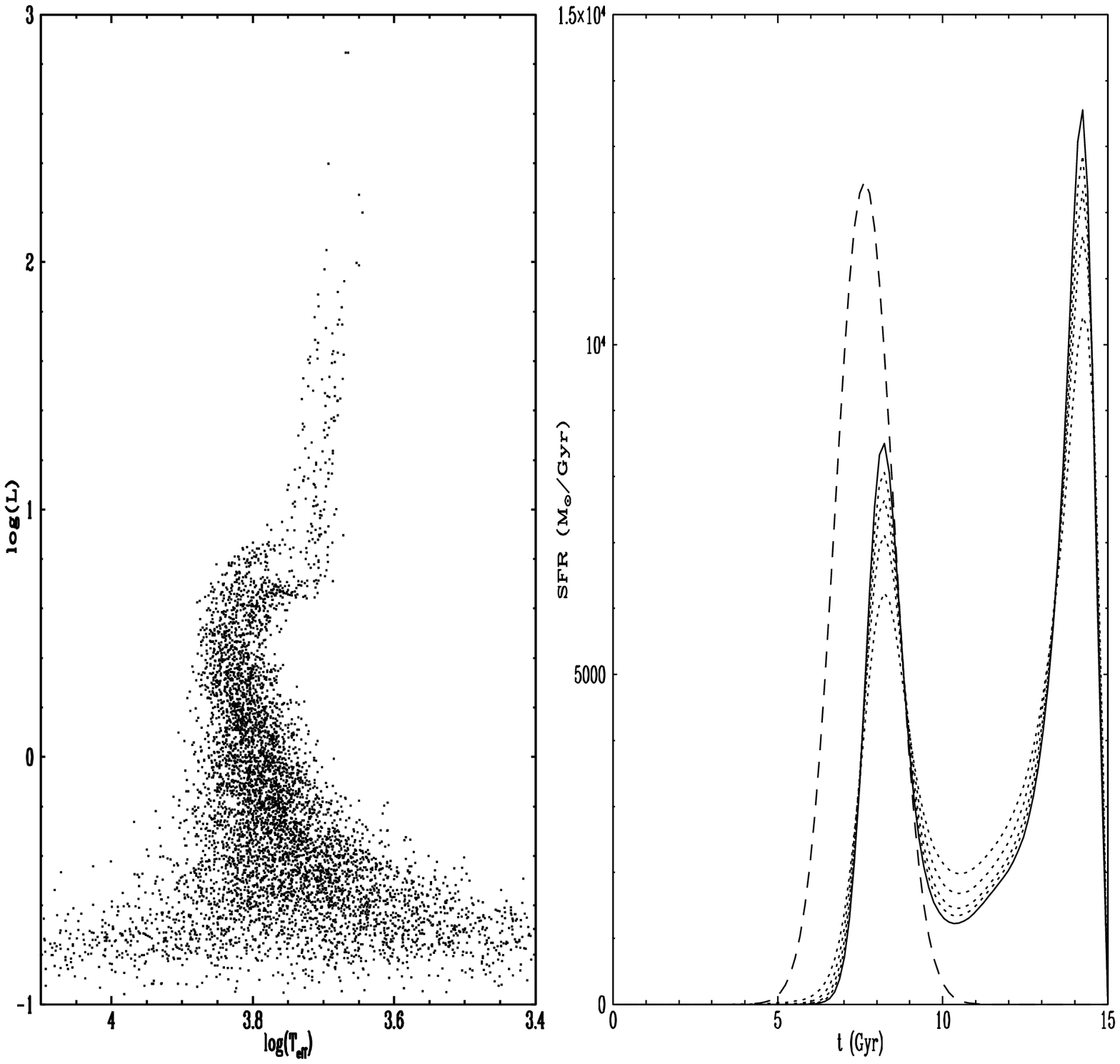,angle=0,width=18.1cm,height=9.0cm}
\@ \textbf{Figure 11.}\hspace{5pt}{
\begin{flushleft}\textbf{Left:} Synthetic HR diagram 
resulting from the sixth input $SFR(t)$, produced using a metallicity 
of $[Fe/H]_{\odot}=-1.7$ for the stars older than 7.5 Gyr and of 
$[Fe/H]_{\odot}=-0.7$ for the stars younger than 7.5 Gyr.
\textbf{Right:} Sixth input $SFR(t)$, dashed line. Also shown
are the  3, 6, 9, 12 and 15 iterations of the inversion method, dotted curves.
The 18th iteration is given by the solid curve, assuming a 
constant metallicity for the entire evolution, which confuses the method.
\end{flushleft}}
\end{figure*}

Figure (11) shows the HR diagram which results from the input $SFR(t)$ of 
Figure (9), with the difference that on this occasion the metallicity
was not constant. In this case we used a metallicity of $[Fe/H]=-1.7$ for
the stars older than 7.5 Gyr, and of $[Fe/H]=-0.7$ for the stars younger than 7.5 Gyr, i.e.,
a crude enrichment history. This is clearly seen in Figure (11), where the two
populations having different metallicities are evident, 
from the width of the RGB. As in all previous cases, 
the noise level was not changed. The result of applying the inversion method
assuming a constant metallicity of $[Fe/H]=-1.7$ is shown in the 
right panel of Figure (11).
The method correctly identifies the half of the $SFR(t)$ with the lower
metallicity; the higher
metallicity population is totally misinterpreted. Actually, the age the inversion
procedure should assign to the high metallicity component is in fact greater than
15 Gyr, which is in contradiction with the fixed boundary condition of
$SFR(15)=0$. This makes the inversion procedure somewhat unstable, which in principle 
can be used to indicate that the 
isochrones being used in the inversion procedure do not correspond to the studied stars.
The two distinct giant branches seen in this HR diagram indicate a difference in the 
metallicities of both populations. If this were introduced as an extra constraint, the 
method would successfully recover the correct $SFR(t)$.

As it might have been expected, uncertainties in the metallicity distort the inference
procedure significantly, making the present method only of any use in cases where the
metallicity is well known, as is the case for some dSph companions of the Galaxy. Other
stellar systems exist where direct spectroscopic measurements can yield an observational
constraint on the metallicity, and consequently make methods such as the one
we have presented here useful. The uncertainties inherent to the determinations of these 
metallicities will result in errors in the answer yielded by the method, in ways similar to those
discussed by Aparicio et al. (1997) or Tolstoy et al. (1998).

However, solving the theoretical problem of
recovering simultaneously the $SFR(t)$ and $Z(t)$ remains a challenge. Variational
calculus methods can be formally extended to the case of recovering several
functions simultaneously, and in a further work we shall present one option through
which the age metallicity degeneracy can be much reduced. Basically, the difficulty
stems from the impossibility of determining uniquely an observed star's age and
metallicity from its observed colour and luminosity. One can in practice pick an
age, and there will be a star of some metallicity with the required properties, which makes
isochrone fitting impractical. Nevertheless, the masses of these stars will be different, 
which provides a way of distinguishing between 
an old, low metallicity isochrone and a young metal rich one, even if the curves are
indistinguishable on the HR diagram. In calculating the full likelihood, one considers
the probability of an observed star belonging to each isochrone. Even if the star is
the same distance from two isochrones of different age, the possibility of it
coming from both will be different, as the density of points along them will depend
on mass. Knowing the relevant IMF therefore furnishes an extra constraint.

\begin{figure*}
\epsfig{file=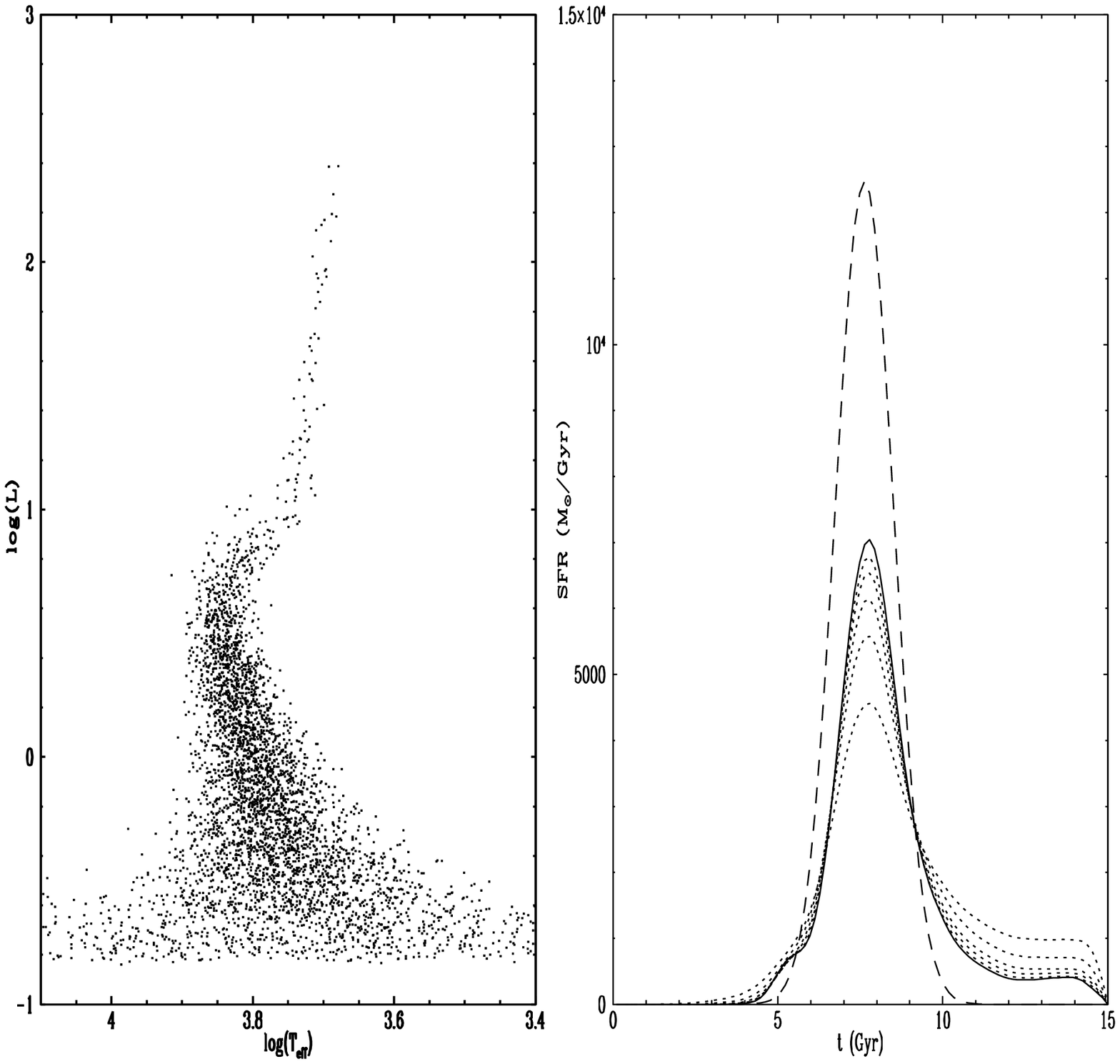,angle=0,width=18.1cm,height=9.0cm}
\@ \textbf{Figure 12.}\hspace{5pt}{
\begin{flushleft}\textbf{Left:} Synthetic HR diagram 
resulting from the sixth input $SFR(t)$, produced using a binary fraction of 0.5.
\textbf{Right:}  
Sixth input $SFR(t)$, dashed line. Also shown
are the  3, 6, 9, 12 and 15 iterations of the inversion method, dotted curves.
The 18th iteration is given by the solid curve, assuming a binary 
fraction of 0, which results in a normalization error.
The slight broadening of the MS is interpreted by the method as a small older
component.
\end{flushleft}}
\end{figure*}

\subsection{Binaries mismatch}

As a final variation we consider the effects a non-zero binary fraction would produce,
which is shown in Figure (12). Figure (12) shows the HR diagram which results from 
the $SFR(t)$ of the previous tests, with the same IMF and metallicity of Figure (9), but
with the inclusion of a binary fraction of 0.5. Half of the stars generated had a secondary
companion picked from the same IMF. The luminosity of the resulting
binary is given by the sum of the luminosities of the two components,
and its combined effective temperature through the Stefan-Boltzmann law. In the
current observations of dSph galaxies and other similarly crowded
fields the main contribution to the ``binary'' population comes not
from physical binaries, but from observational confusion. Attempting to
model this effect we picked the secondary star from the same IMF as
the primary one (e.g. see Kroupa et al. (1993) for a discussion of
binary confusion in observations). We took the value of 0.5
for the binary fraction as a representative number from Kroupa et al. (1993). 

As can be seen from comparing Figures (12) and (9), the main result of having
included a large binary fraction is a reduction in the total number of stars;
the morphology of the HR diagram was not significantly affected. This last
effect is due to the fact that the offset between the single star and the
binary star main sequences is comparable to the noise in that region, producing only
a slightly broadened main sequence. The effects of binaries in other parts of the
diagram are negligible, as the addition of a main sequence star to a giant 
does not affect the observed properties of the giant, and 
the odds of getting a binary giant are slim. The results of 
the inversion method are shown in the right panel of Figure (12), where the dashed line shows the
input $SFR(t)$, and the dotted and solid curves the first 18 iterations of the method, every 3.
The convergence of the method is not affected, and proceeds quite rapidly.
It can be seen that the method accurately identifies the age, duration and structure
of the input burst, although with a normalization error which results from the reduction
in the total number of stars seen. A further slight discrepancy between the input
$SFR(t)$ and the recovered one appears at old ages, as the method confuses the broadening
in the main sequence for a minor, extended age population. As with the errors in the 
IMF, having neglected the effects of binaries affects mostly the normalization
of the recovered $SFR(t)$, distorting the general shape only slightly. This is
encouraging as the absolute normalization will in any case have to be supplied
externally through global $M/L$ ratios or other observations, when studying large extended
systems such as dSph galaxies for which HR diagrams, covering the complete galaxy
 do not at the moment exist.

\section{Conclusions}

We can summarize our results as follows:

1) We have introduced a variational calculus scheme for solving maximum
likelihood problems, and tested it successfully in the particular case of
inverting HR diagrams.

2) Assuming a known IMF and metallicity we have presented a totally
non-parametric method for
inverting HR diagrams which yields good results when recovering stellar
populations younger than 10 Gyr, with data quality similar to
those attained in current HST observations of dSph galaxies.
Populations older than 10 Gyr can only be recovered equally well from HR
diagrams with much reduced observational errors.

3) Uncertainties in the IMF and binary fractions result in normalization
errors on the total $SFR(t)$. Given the existence of an age-metallicity
degeneracy on the colours and magnitudes of stars, an error in the assumed
metallicity results in a seriously mistaken $SFR(t)$. This makes the version of 
the variational calculus approach we present here only useful in cases where the
metallicity of the stars is known.

\section*{Acknowledgments}

The work of X. Hernandez was partly supported by a DGAPA-UNAM grant. The authors wish to
thank Rafael Rosales for his help in developing the statistical model presented here and
Ricardo Carretero for help with the convergence of the method.

\begin{figure*}
\epsfig{file=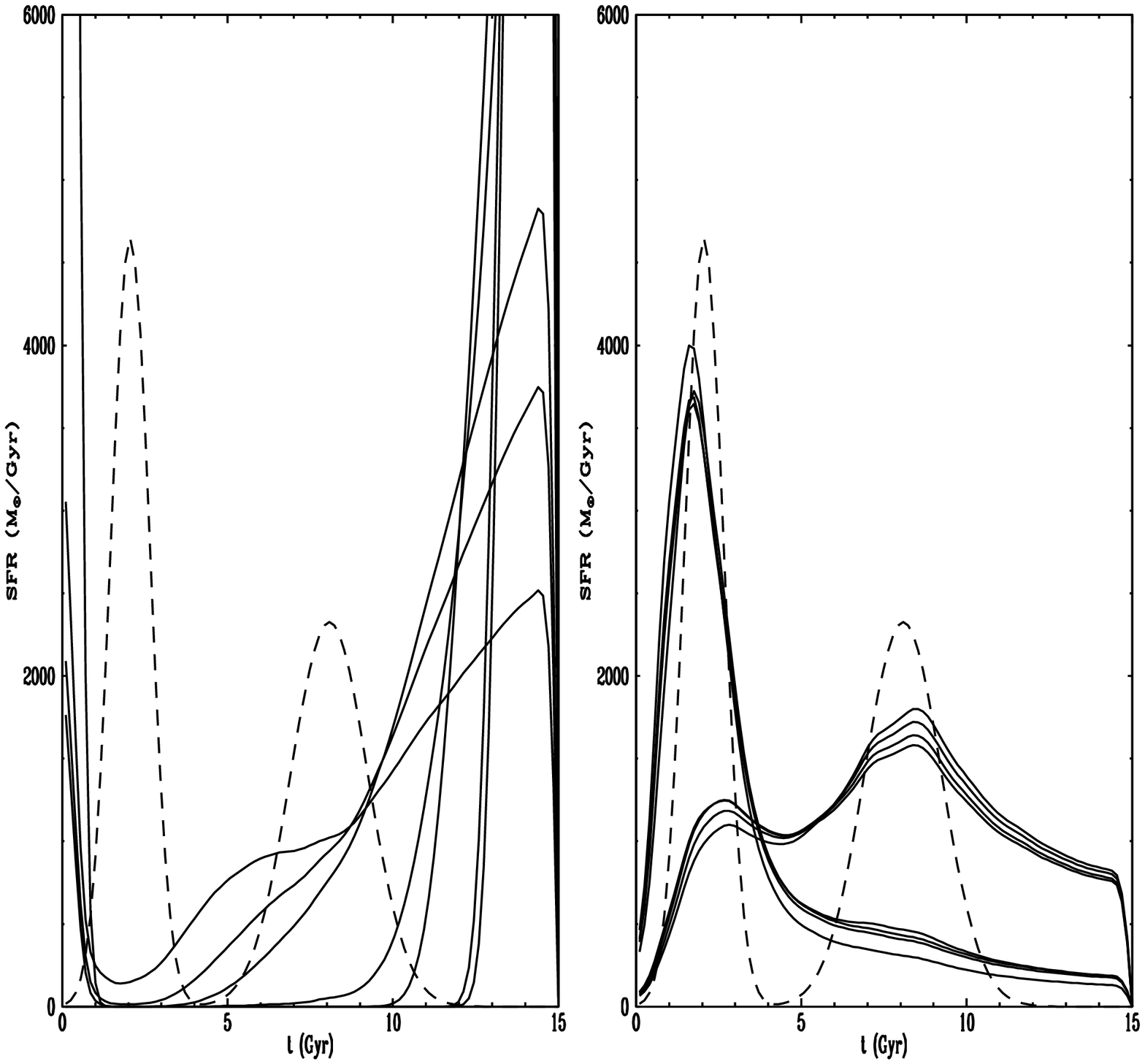,angle=0,width=18.1cm,height=9.0cm}
\@ \textbf{Figure A1.}\hspace{5pt}{
\begin{flushleft}\textbf{Left:} Test input $SFR(t)$, dashed line. Also shown are the first 4
iterations of the method as implied by equation (8), which diverges to a minimum of the likelihood.
\textbf{Right:}  
Test input $SFR(t)$, dashed line. The solid curves show
the  1, 2, 3, 4, 5, 8, 12 and 13 iterations of the inversion method, after having inverted
the sign of the derivative of the solution of equation (8).  The method now oscillates
between two families of curves, each of which diverges.
\end{flushleft}}
\end{figure*}

\begin{figure*}
\epsfig{file=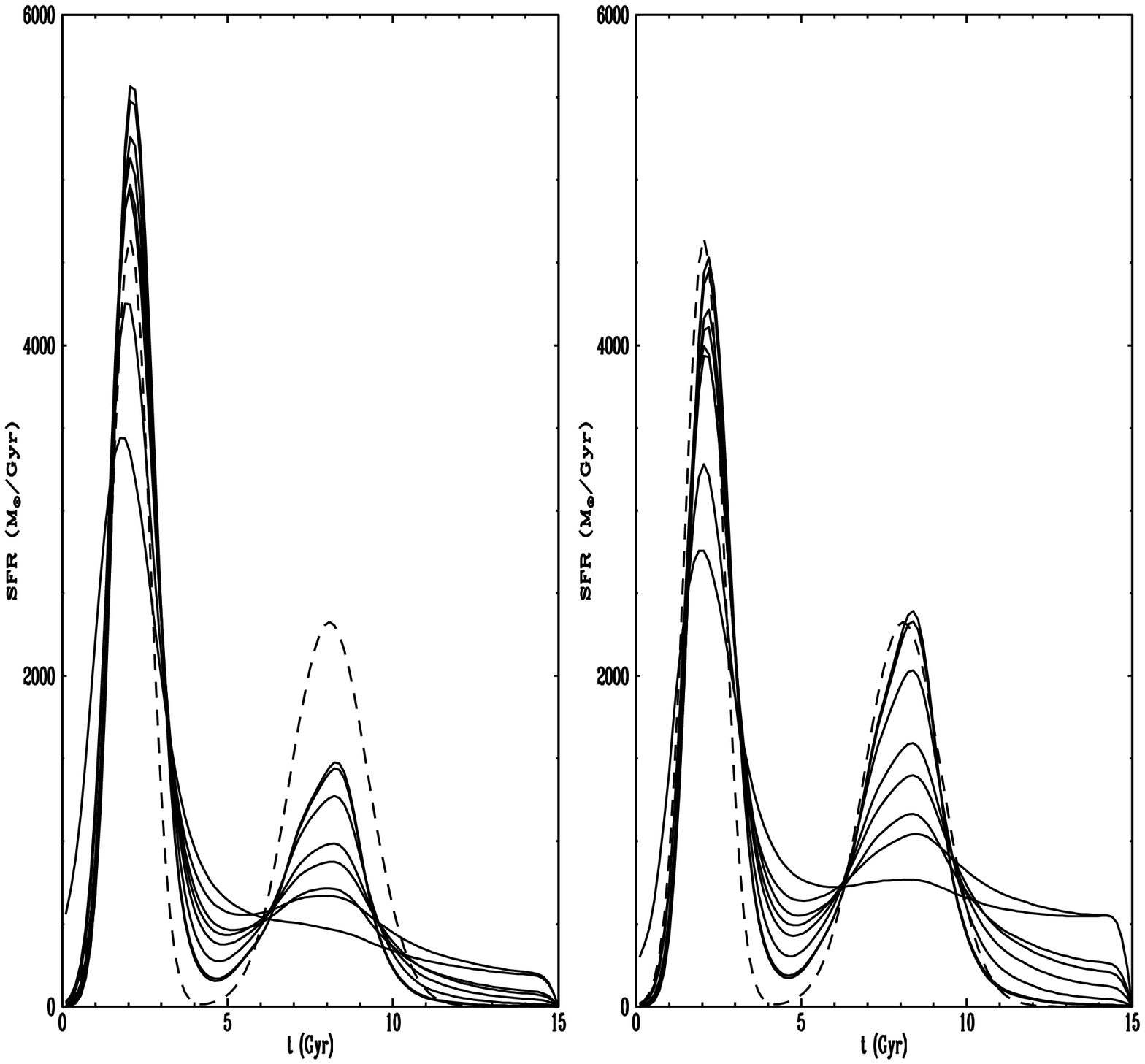,angle=0,width=18.1cm,height=9.0cm}
\@ \textbf{Figure A2.}\hspace{5pt}{
\begin{flushleft}\textbf{Left:} Test input $SFR(t)$, dashed line. The solid curves again show
the  1, 2, 3, 4, 5, 8, 12 and 13 iterations of the inversion method, once an averaging across
iterations is introduced on the coefficient of Equation (8) which oscillates. Convergence
is now rapid, and only a preponderance of the younger populations remains.
\textbf{Right:}  
Test input $SFR(t)$, dashed line. The solid curves show
the  1, 2, 3, 4, 5, 8, 12 and 13 iterations of the inversion method, after correcting for
the differential incompleteness in the HR diagram due to the absence of stars more massive than
the tip of the RGB, which decreases with time.
\end{flushleft}}
\end{figure*}

\appendix

\section{numerical implementation and convergence}

In section 2 we proposed a variational calculus approach to solving directly for the
maximum likelihood $SFR(t)$, which yielded:

\begin{equation}
{d^2 Y(t)\over dt^2} C1(t)=-{dY(t)\over dt}C2(t)
\end{equation}

where 
$$
C1(t)=\sum_{i=1}^{n} \left( G_{i}(t) \over I(i)\right), C2(t)=\sum_{i=1}^{n}
 \left( dG_{i}/dt \over I(i)\right)
$$
and
$$
I(i)=\int_{t_0}^{t_1} SFR(t) G_{i}(t) dt
$$

The first step is to evaluate $G_{i}(t)$ from the position of the modeled stars
for each of the values of t being considered, 100 in this case.
$dG_{i}/dt$ is then calculated using a quadratic interpolation on $G_{i}(t)$.
We next evaluate the coefficients in equation~(8) by assuming
$I_{0}(i)=1 \hspace{10 pt} \forall i$. Setting the condition
$Y'(0)=0.0001$, we solve equation~(8) using
a finite differences scheme, to obtain $SFR_{0}(t)$. This is then used
to recalculate the integral coefficients for each star, $I_{1}(i)$,
which completes the cycle of each iteration.
Other initialization procedures are of course possible, but this one
is the one which appears natural to the problem, as it gives a
$SFR_{0}(t)$ which is determined by the likelihood matrix.
$SFR(t)$ is then recovered
from $Y'(t)$ and normalized through the total number of stars. The results
of this iterative scheme are shown by the solid lines in the left panel of 
Figure (A1), which also shows as a dotted line the input $SFR(t)$. This is the same
case as shown in Figure (1).

As the left panel of Figure (A1) shows, the method gives a solution which is actually 
the opposite of the input $SFR(t)$, and which rapidly diverges and oscillates. This result is not
altogether surprising, as variational calculus methods yield solutions which imply
stationary values of a functional, not necessarily a maximum. In this case, the method
is tending to a {\it minimum} of the likelihood. To overcome this problem we introduce
the conjecture that the maximum likelihood $SFR(t)$ will have a time derivative of the
opposite sign to that of the one which minimizes the likelihood, at all points. 
This assumption translates into changing the sign of C2(t).

The result of the method is now shown in the right panel of Figure (A1) which gives 
the 1,2,3,4,5,8,12 and 13 iterations, as in the left panel. In this case, the method correctly identifies
the ages of the populations present, but actually oscillates, and each branch still slowly
diverges. This behaviour is not uncommon in dynamical systems, and there was no guarantee
that the simplest implementation would work. Inspecting the behaviour of the C1(t) and
C2(t) coefficients we identified C2(t) as the source of the oscillations. We therefore
introduced an averaging from one iteration to the next on the value of C2(t). As in each 
branch one population grows as the other tends to zero, we kept twice the average of
C2(t) at the present iteration and the previous one.

The result of the method at this stage can be seen in the left panel of figure (A2).
The dotted curve is again the input $SFR(t)$, and the solid ones show the same iterations
as in the previous case. The method now converges steadily into the two populations
present, correctly excluding populations not present, although a tilt towards younger
ages is apparent. We interpret this tilt as caused by the differential incompleteness
of the HR diagram as a function of time. As our isochrones stop at the tip of
the RGB, at a star of a zero age mass which decreases for older isochrones,
the older populations are missing an increasing proportion of their stars. 
To obtain the $SFR(t)$ in the mass range $0.08 M_{\odot} - \infty$ from our
inferred $SFR(t)$ which sees stars only between $0.6 M_{\odot}$ and the tip
of the RGB we multiply the $SFR(t)$ by a minor correction factor to
compensate for this incompleteness, given by integrating equation~(9)
out to the tip of the RGB at every t, and comparing with the result at t=0. 
This final correction is introduced only when plotting the results, and is not
included in the iterative procedure. In dealing with real data it will be necessary to
remove any stars associated with evolutionary phases not included in the isochrones
with which those stars are compared.

The final results for this particular example are shown in the right panel of 
Figure (A2), where the solid curves show the same iterations as the previous 2 cases.
The method now tends strongly to the input $SFR(t)$, to a level limited only
by the amount of noise present in the modeled stars, as explored in section (3).
We realize that the final method is not a rigorous implementation of the mathematical
hypothesis, for which reason we have tested the validity and precision experimentally
with a very large number of synthetic HR diagrams. The results where highly accurate
in all cases, a fraction of which are detailed in sections (3) and (4).

\end{document}